\documentclass[]{aastex7}

\newcommand{\mafont}{
  \bfseries
  \color{blue}
}

\newcommand{\orcca}{ORCCA} 

\usepackage{hyperref}
\usepackage{soul}
\DeclareTextFontCommand{\ma}{\mafont}
\usepackage[normalem]{ulem}
\newcommand\maout{\bgroup\markoverwith{\textcolor{blue}{\rule[.5ex]{2pt}{1.0pt}}}\ULon}

\usepackage{hyperref}
\begin{document}

\title{Observation and Modeling of Shear Evolution of Post-reconnection Flare Loops}

\author{Drake Osaben}
\affiliation{Department of Physics,
Montana State University,
Bozeman, MT 59715, USA}
\email{drakeosaben@montana.edu} 

\author{Jiong Qiu}
\affiliation{Department of Physics,
Montana State University,
Bozeman, MT 59715, USA}
\email{qiu@montana.edu} 

\author{Dana W. Longcope}
\affiliation{Department of Physics,
Montana State University,
Bozeman, MT 59715, USA}
\email{longcope@montana.edu} 


\begin{abstract} 

A solar flare releases magnetic energy by reconnecting field lines across a current sheet, thereby allowing their relaxation to a lower energy state.  The maximum possible energy is released if all field lines relax to a current-free (potential) state.  The progress of a flare's reconnection is often measured as the angle-complement between the observed post-reconnection flare loops and the polarity inversion line of the photospheric magnetic field: shear angle.  Many observations have shown strong-to-weak shear evolution over the course of a flare.  A field line's shear angle is, however, an imperfect measure of its relaxation.  We develop a new technique for observationally inferring the three-dimensional structure of post-reconnection field lines, including their local twist, $\alpha$, which will vanish for potential fields.  Our method fits loops in EUV images to extrapolations subject to constraints such as matching the feet of model field lines to observed flare ribbons.  We apply the new method to an eruptive two-ribbon flare (SOL2014-12-18T22) which exhibits strong-to-weak shear-angle evolution. We find that, as the flare progresses, $\alpha$ decreases in post reconnection loops anchored to newly brightened ribbons. Our study demonstrates that post-reconnection magnetic field is neither potential nor linear force-free. The method quantifies, for the first time, the time-history of a flare's energetic relaxation.  It also quantifies the increasing height of subsequently reconnected field, and the time delay between reconnection forming a flare loop and its appearance in EUV passbands.  These results promise to enable improvements in both magnetic modeling and hydrodynamic modeling of flares.


\end{abstract}

\keywords{}

\section{Introduction} \label{sec:intro}

A solar flare releases free magnetic energy by magnetic reconnection that relaxes magnetic field to a lower-energy state. The configuration of a particular type of flares, eruptive two-ribbon flares, has been depicted by the standard model, well known as the Carmichael-Sturrock-Hirayama-Kopp-Pneuman (CSHKP) model \citep{Carmichael1964, Sturrock1966, Hirayama1974, Kopp1976}. The model explains a number of key observational signatures: magnetic reconnection occurs in the corona between a set of anti-parallel magnetic field lines across a vertical current sheet, and forms an arcade of flare loops in the corona with their feet anchored in a pair of flare ribbons in the chromosphere on the two sides of the polarity inversion line (PIL) separating positive and negative magnetic flux. Most of the reconnection released energy 
rapidly deposits at the feet of the flare loops, producing bright flare ribbons. The impulsive energy deposition in the chromosphere  
drives mass upflows \citep[chromospheric evaporation;][]{Antonucci1982, Fisher1987} into the coronal loops, which then become prominently visible in the soft X-ray (SXR)  and then Extreme UltraViolet (EUV) light as they cool from a few tens to a few million kelvin. { The pair of flare ribbons in the chromosphere are observed in visible, ultraviolet (UV), and every now and then hard X-ray (HXR), emissions.  Observations often show that the two flare ribbons move away from the PIL during the flare \citep{Svestka1980, Kitahara1990, Sakao1994, Fletcher2001, Isobe2002, Asai2004, Qiu2002, Temmer2007, Liu2009a, Yang2011, Qiu2017, Kazachenko2017, Hinterreiter2018}. The apparent ribbon motion is interpreted as a signature} of progressive magnetic reconnection proceeding continuously and forming taller, and presumably longer, loops with their conjugate feet at progressively larger distance from each other.


Despite its success in capturing the global signatures of flare evolution, the standard %
model cannot account for many other properties of flare reconnection. Flares tend to occur in sheared magnetic fields \citep{Schrijver2008, Bobra2015, Patsourakos20}, where reconnection must take place between a pair of field lines that are not anti-parallel (i.e., component reconnection). Therefore, post-reconnection field lines will carry a non-vanishing component along the direction of the current, which is also the direction of the PIL in a 2D or 2.5D\footnote{2.5D refers to the configuration with a symmetry direction over which quantities do not vary but along which vectors may have non-vanishing components.} configuration. Such a feature has been characterized observationally by the shear angle $\theta$ of post-reconnection flare loops (PRFLs), defined as the complement of the angle between a PRFL projected on the solar surface and the PIL of the vertical component of the photospheric magnetic field. Observations of many eruptive flares have shown that earlier formed flare loops, when viewed top-down in disk observations, appear to have a larger shear angle, than later formed loops. This is known as the ``strong-to-weak'' shear evolution of PRFLs \citep{Aschwanden2001}.

The apparent shear angle of PRFLs has been measured in a number of studies { over} two decades. Before the Solar Dynamics Observatory \citep[SDO;][]{Pesnell2012} was launched, the apparent shear angle $\theta_{ft}$ was estimated by measuring the angle between the PIL of the line-of-sight component of the photospheric magnetic field, approximated as a straight line, and another straight line connecting a pair of flare kernels or centroids of flare ribbons observed in optical \citep{Ji2006}, UV \citep{Su2006, Su2007}, or HXR \citep{Sakao1994, Bogachev2005, Grigis2005, Yang2009, Liu2009b, Inglis2013, Zimovets2020} emissions. These methods primarily assume the magnetic connectivity between the most prominent pair of flare kernels or centroids of flare ribbons. The strong-to-weak shear evolution was often, though not always, associated with the parallel-to-perpendicular apparent motion of flare ribbons or kernels with respect to the PIL, which is often curved rather than straight. 
\citet{Qiu2009, Qiu2010, Qiu2017, Qiu2022} therefore defined a shear index $ \mathcal{S} \equiv \Delta \langle l_{||}\rangle /\Sigma \langle l_{\perp}\rangle $, where $\langle l_{||} \rangle$ is the distance of the centroid of flare ribbon fronts along the curved PIL, $\langle l_{\perp}\rangle $ is the mean distance of the centroid perpendicular to the PIL, and $\Delta$ and $\Sigma$ refers to the difference and sum, respectively, of the measurements for ribbons in the positive and negative magnetic fields. The shear index is another proxy of the mean shear of the post reconnection flare arcade with respect to the curved PIL, and measurements in several two-ribbon flares reported in these papers also confirmed the strong-to-weak shear evolution with $\mathcal{S}$ varying from up to 5 at the start of the flare to below 2 or 1 when the flare HXR emission at $> 20$~keV peaks.  

Observational studies in recent years have taken advantage of PRFLs directly observed by SDO's Atmosphere Imaging Assembly \citep[AIA;][]{Lemen2012}. \citet{Li2019} measured the shear $\theta_{lp}$ between several PRFLs with the PIL of the line-of-sight photospheric magnetic field in about 20 flares. \citet{Qiu2017} measured the shear in a slightly different way, where $\theta_{lp}$ is derived from the angle of a set of observed loops with the PIL of the extrapolated potential magnetic field at an assumed height of the loops. 
Most recently, \citet{Qiu2023} have identified up to 3,000 PRFLs from AIA observation in two EUV passbands characterized by plasmas of $\sim$1~MK. The general trend of the $\theta_{lp}$ also exhibits the strong-to-weak shear evolution, and is consistent with shear index evolution derived from the ribbons.

The strong-to-weak shear evolution, and the apparent parallel motion of flare kernels or ribbons, are clear observational signatures of three-dimensional magnetic reconnection \citep{Aulanier2012, Janvier2013}. 
On the other hand, shear angle measurements have been constrained to information in the image plane, and the shear angle $\theta_{ft}$, $\theta_{lp}$, and the shear index $\mathcal{S}$, are, at best, {\em proxies} of some 3D reconnection properties. Furthermore, the measurements are subject to ambiguities introduced by the often complex shape of the photospheric PIL \citep[e.g.][]{Inglis2013}. It is, therefore, difficult to interpret the exact physical meaning of the observed shear evolution. Some idealized numerical models of flare reconnection can reproduce the observed strong-to-weak shear evolution \citep{Dahlin2022}, with the setup of a {\em pre-reconnection} magnetic field that has a smooth distribution in the photosphere, a straight PIL, and more sheared field lines near the PIL. These studies imply that the observed strong-to-weak shear evolution of PRFLs may be reminiscent of the {\em pre-reconnection} field configuration. The observational measurements of shear evolution, however, must be associated with {\em post-reconnection} configurations. It remains to be understood how reconnection relaxes the pre-reconnection field to generate the observed post-reconnection signatures. The 3D magnetic field in active regions, both {\em before} and {\em after} flare reconnection, is more complex than idealized MHD models \citep[e.g.][]{Sun2012}. The post-reconnection field describing the observed PRFLs has not been modeled, 
yet the knowledge of which is crucial for understanding relaxation of magnetic field by flare reconnection.



In this paper, we present a novel experiment that constructs the 3D magnetic field of PRFLs of an M6.9 eruptive two-ribbon flare to understand the physical nature of the strong-to-weak shear evolution. The model field is an optimized fit to the observed PRFLs, with the constraint that field lines must be anchored to observed flare ribbons. In the following sections, we present an overview of the flare and 3,000 PRFLs identified from EUV images (\S\ref{sec:overview}). We describe in \S\ref{sec:method} the data-constrained method, the Optimized Ribbon-Constrained Constant-$\alpha$ (ORCCA) fitting method, to model these 3,000 PRFLs. Our results show that PRFLs formed earlier deviate more from the potential field than later formed PRFLs (\S\ref{sec:results}). The physical implication of these results are discussed in \S\ref{sec:conclusion}.

\section{Overview of Observations}
\label{sec:overview}

The SOL2014-12-18 M6.9 flare was accompanied by a Coronal Mass Ejection (CME). Comprehensive descriptions of the flare observed by multiple telescopes were provided in \citet{Joshi2017, Qiu2023}, and data-driven MHD simulations of the eruption were conducted by \citet{Prasad2023}. For this event, a large number of post-reconnection flare loops are visible in the EUV images in the 304~\AA\ and 171~\AA\ passbands of AIA throughout the flare evolution. EUV images in other EUV bands are largely saturated, and are therefore not analyzed in this study.

\begin{figure}     
    \centering
    {{\includegraphics[width=\textwidth]{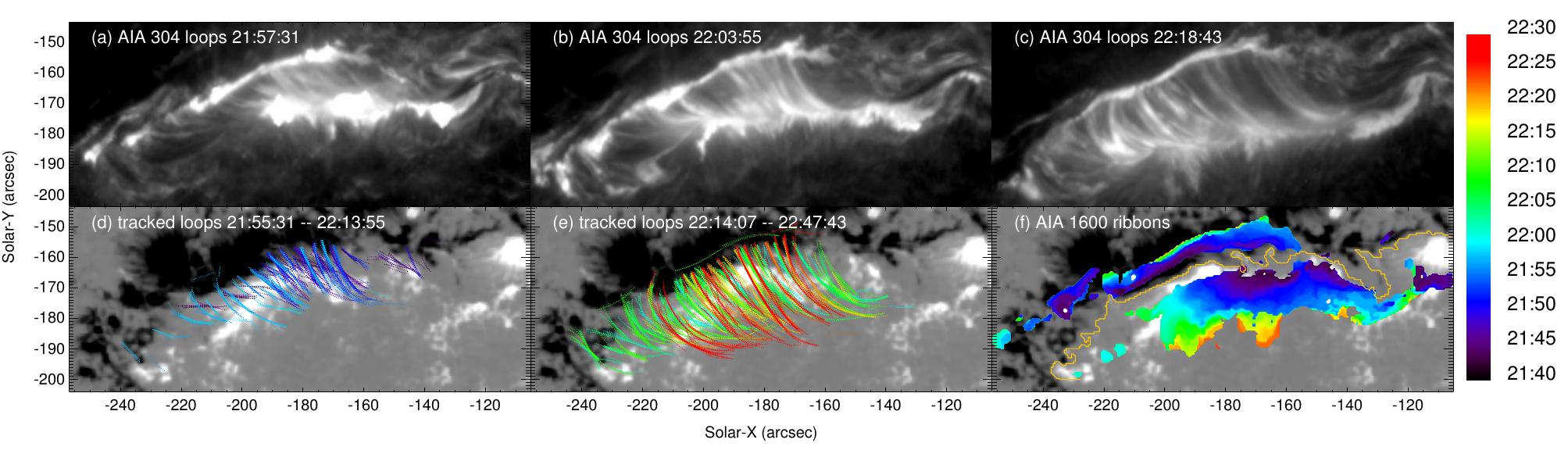} }}%
    \caption{Top: post-reconnection flare loops observed in the EUV 304~\AA\ images by AIA at different times. Bottom: (d-e) PRFLs identified from the AIA 304~\AA\ images using the OCCULT code,  superimposed on a magnetogram of the photospheric radial magnetic field obtained by HMI at 21:00~UT. The colors indicate the times when the loops were observed in the AIA 304~\AA\ images minus 15 minutes. (f) Evolution of flare ribbons observed in the AIA 1600~\AA\ passband, with the colors of the contour indicating the time the ribbon pixels are brightened, as denoted in the color bar in the right. The orange curve outlines the PIL of the photospheric radial magnetic field (gray scale). All images in this figure have been rotated to 21:00 UT, and the helioprojective coordinates in the figures reflect this reference time.}%
    \label{fig:overview}%
\end{figure}

The PRFLs can be identified using the Oriented Coronal CUrved Loop Tracing algorithm \citet[OCCULT;][]{Aschwanden2010}. 
The code 
applies a high-pass filter to identify bright pixels against the background, and traces neighboring bright pixels that form an arc, which is considered to be a loop, or part of a loop. 
\citet{Qiu2023} applied the OCCULT code to EUV observations of this flare. In the 304~\AA\ passband, 1980 loops, or parts of loops, were identified between 21:55 – 22:47~UT using 201 images of 12~s cadence. Images in the 171~\AA\ passband are saturated in the early phase of the flare, so the analysis was applied to 98 images (of 24~s cadence) between 22:04 - 22:46~UT, yielding 933 loops. Figure~\ref{fig:overview} shows the PRFLs observed in the 304~\AA\ images (top panels), and also the PRFLs identified from these images (bottom panel), superimposed on a magnetogram of the photospheric radial magnetic field obtained by the Helioseismic and Magnetic Imager \citep[HMI;][]{Schou2012}. Note that a single PRFL may be identified in multiple frames in a EUV passband since loop emission typically persists for a few minutes in the passband \citep{Aschwanden2001, Qiu2016}. Also note that a single PRFL in one EUV image may appear as several disconnected parts or sections, and therefore is identified as several ``loops". As a result, the OCCULT code has identified more than a thousand loop-like structures in one passband during the flare evolution. For simplicity, we refer to all these structures as ``loops" in the { following} text. \citet{Qiu2023} analyzed and modeled evolution of these loops identified in the 304~\AA\ and 171~\AA\ passbands, and found that loops identified in these two passbands form at nearly the same temperature, $\le$ 1~MK.

Figure~\ref{fig:overview}f also shows the evolution of flare ribbon brightening observed in the AIA 1600~\AA\ images. The color indicates the onset time of the ribbon brightening, defined as when a ribbon pixel becomes 6 times brighter than the pre-flare quiescent sun and continues to be bright for more than 4 minutes \citep{Qiu2023, Qiu2025}. We consider that this is approximately the time when reconnection occurs in the corona forming a PRFL anchored at the UV-brightened pixel. Figure~\ref{fig:overview}f therefore shows the evolution of flare reconnection forming the PRFL arcade. Seen from the figure, the identified PRFLs are apparently anchored at the two ribbons in magnetic fields of opposite signs. On the other hand, PRFLs become visible more than 10 minutes after the ribbon brightening, reflecting the timescales of plasma heating and then cooling to $\sim$1~MK to be observed in these passbands.


Seen from the figure, earlier formed PRFLs (violet and blue) appear to be more inclined toward the PIL of the photospheric magnetogram than later formed PRFLs (green, orange, and red), suggesting the strong-to-weak shear evolution as reported in many previous studies. For this flare, the strong-to-weak shear evolution was inferred from the evolution of flare ribbons \citep{Qiu2022}, and was then confirmed with $\theta_{lp}$ measured with $\sim$3,000 PRFLs observed in AIA 304 and 171 passbands characterizing plasma emission at 1~MK \citep{Qiu2023}. In this paper, we will model the three-dimensional magnetic structure of these loops to better understand the nature of the shear evolution.

\section{Modeling the Magnetic Field of Post-Reconnection Flare Loops} 
\label{sec:method}


In solar flares, magnetic reconnection forms PRFLs, which then relax on Alfv\'enic timescales, typically of a few seconds, under magnetic Lorentz force.  Plasmas inside PRFLs are heated to more than 10 MK and then cool down to $\sim$1~MK in 10$^{0-1}$ minutes. Therefore, we consider PRFLs observed in the EUV passbands characteristics of plasma temperature $\sim$1~MK to be nearly force free, satisfying $\nabla \times {\bf B} = \alpha {\bf B},$ where $\alpha$ is the force-free parameter. Conventionally, the 3D magnetic field is constructed using an extrapolation method that solves the force-free equation and Maxwell equations, subject to the boundary condition provided by the photospheric magnetograms \citep[see review in][]{Wiegelmann2021}. In the case of a linear force free field (LFFF), the force free parameter $\alpha$ is a constant, and the normal component of the magnetic field in the photosphere is used as the boundary condition. For a non-linear force-free field (NLFFF) with varying $\alpha$, all three components of the vector magnetic field measured in the photosphere are used as the boundary condition. 

In this study, we find that the observed PRFLs cannot be well described by a single value of $\alpha$, which means that the field associated with PRFLs is not a constant-$\alpha$ force-free field (i.e.\ LFFF).  Our intent, however, is to characterize PRFLs immediately following their relaxation in the midst of a flare.  The ongoing flare compromises the vector field measurements making their use in a full NLFFF extrapolation problematic. Moreover, the overall field is rapidly evolving during the flare, and is not an equilibrium.  While each PRFL is assumed to be fully relaxed, and thus in equilibrium, the field nearby 
is undergoing continuing reconnection and retraction, and will not be an equilibrium.
Therefore, we develop a different approach to approximate the 3D force-free equilibrium, with spatially varying $\alpha$, using several kinds of observations, rather than the magnetograms alone, to constrain the model.  We hereafter refer to this method as Optimized Ribbon-Constrained Constant-$\alpha$ fitting (\orcca\ fitting).

\orcca\ fits each PRFL, found from an EUV image, 
to a field line from a particular LFFF, with its prescribed value of $\alpha$.  The fitting yields different values of $\alpha$ for each field line, so the overall field is not itself a force-free equilibrium. \cite{Malanushenko2009} developed and applied a similar method and verified that the collection of field lines, with varying $\alpha$, can approximate a genuine NLFFF with reasonable accuracy. In the present work we further require model field lines to be anchored in flare ribbons,  which are foot-points of PRFLs, observed in the UV 1600~\AA\ by AIA.  In this way \orcca\ takes advantage of the information available from both coronal and chromosphere images and photopsheric magnetogram.


\subsection{The \orcca\ Fitting Method}
\label{subsec:model}


We extrapolate the LFFFs from HMI's cylindrical equal-area (CEA) map \citep{Bobra2014} of the radial component of the magnetic field prior to the flare. The extrapolation is performed in tangent plane coordinates, where the $(x,y)$ plane is tangent to the solar surface and $z$ is the height above that plane.  Radial field measurements from HMI are mapped onto the tangent plane (i.e. $z=0$ surface) and used as a lower boundary to extrapolate a LFFF.  Extrapolations are done for each of 50 different $\alpha$ values ranging uniformly over $0\,{\rm Mm}^{-1}\le\alpha\le 0.2\,{\rm Mm}^{-1}$.  { The range of $\alpha$ values used in this experiment is chosen by visually comparing the extrapolated field lines with subsets of PRFLs observed during various stages of the flare. }

\begin{figure}
    \centering
    
    \label{fig:Top:}
    \includegraphics[width=.52\linewidth]{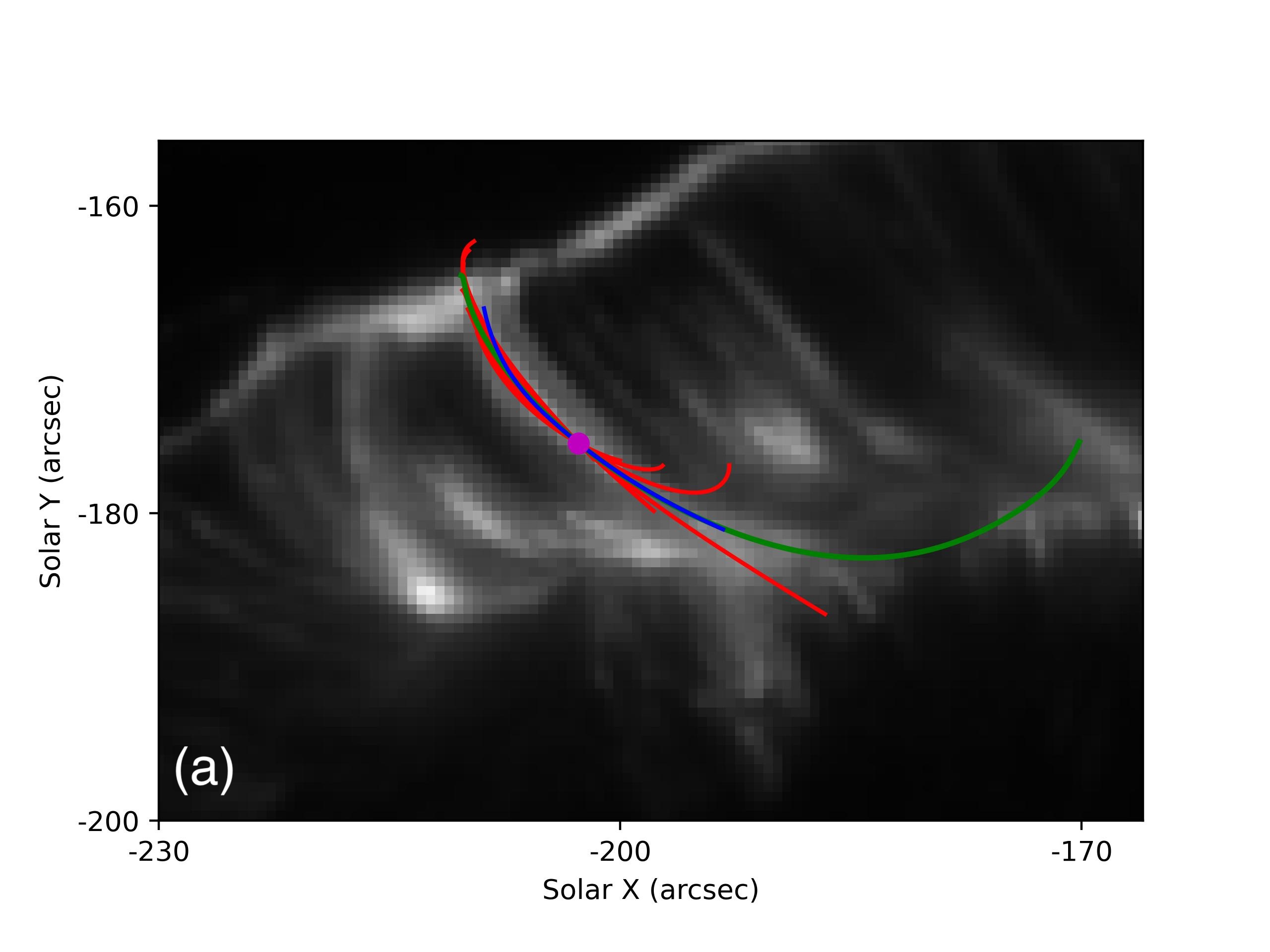}
    \includegraphics[width=.43\linewidth]{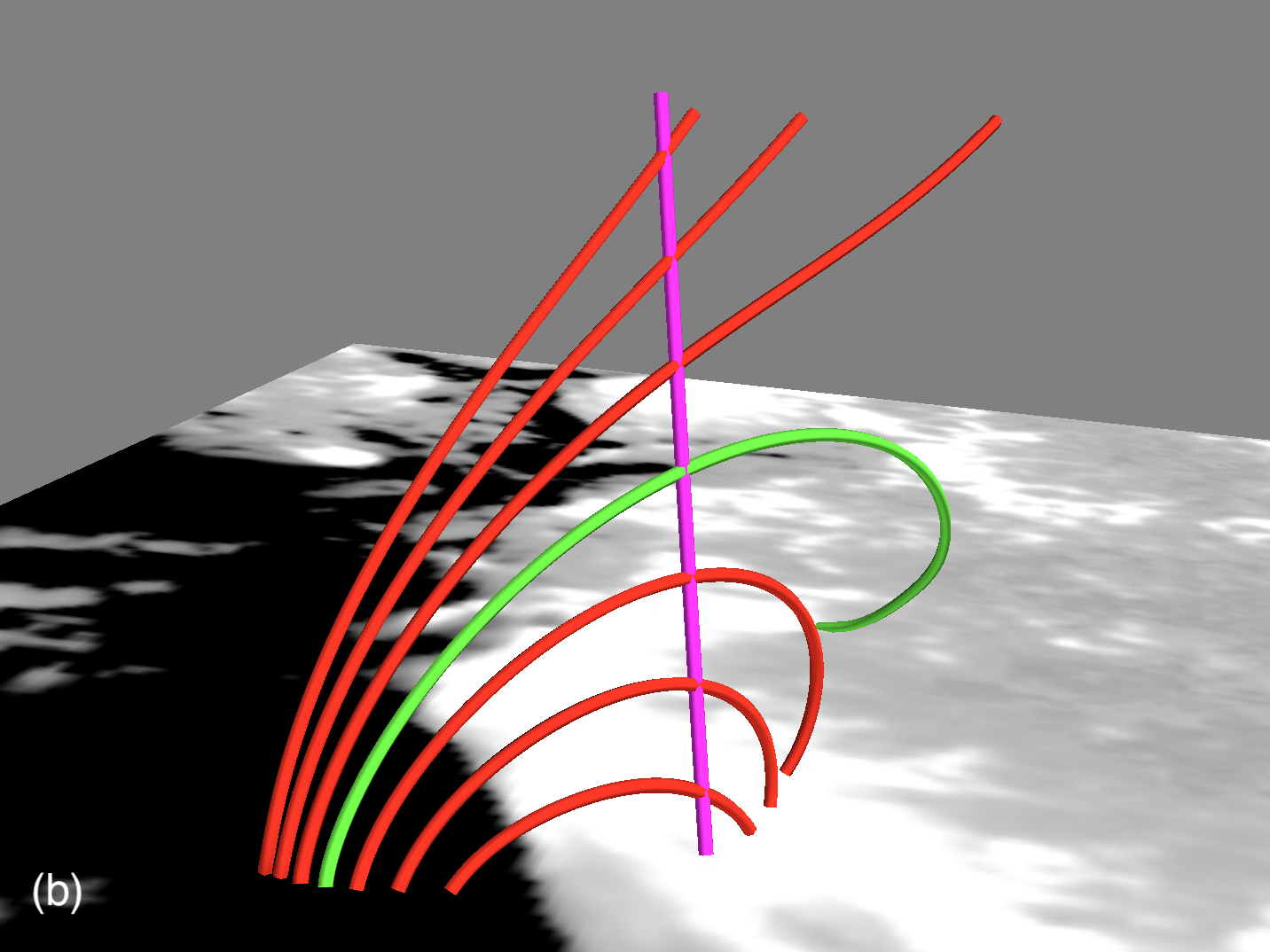}
    
    \caption{(a) A sample of field lines traced at various heights for a given $\alpha$, overlaid on top of an AIA 304 \AA\ image that the (blue) observed loop is identified in. (b): A side view of the same sample of field lines traced with a CEA magnetogram placed at the surface $z=0$, saturated at $\pm$300~G. In both panels, magenta color indicates the observed loop's center pixel, red curves show the sample of model field lines traced at various heights, and the green curve shows the modeled loop that is the best fit to observation.}

\label{fig:model}
\end{figure}

We use each of the 50 fields to generate magnetic field lines in search of a best fit to a particular PRFL.  The pixel at the center of the observed PRFL defines a line-of-sight, which is mapped into the tangent plane space, as illustrated by the magenta line in Figure~\ref{fig:model}b.  { The center pixel of an observed loop is typically a few pixels away from the PIL, because an observed loop is only part of the full loop, and the PIL is closer to the northern ribbon due to asymmetric magnetic field distribution (see Figure~\ref{fig:overview}f).} For a given value of $\alpha$, a single field line is traced in both directions from a point $H$ along this line.  A set of field lines are traced from 40 different values of $H$ ranging uniformly over $0<H\le 20$ Mm; several are illustrated by the red and green lines in Figure~\ref{fig:model}b.  These are then mapped back to the plane-of-the sky, as shown in Figure~\ref{fig:model}a, where they may be successively compared to the observed PRFL, shown as a blue curve in this figure.



Tracing field lines from 40 different heights, $H$, for each of 50 different $\alpha$ values, results in 2000 projections which must be compared to the single PRFL.  We deem the best fit to be that field line which minimizes an error metric with its PRFL; we present the details of the error metric below.  \cite{Malanushenko2009} followed the same approach, but found that the global minimum over the full $(\alpha,H)$ space was not always the most physically plausible field line.  They were fitting loops in quiescent active regions, and considered the full set of traced field lines.  Considering PRFLs in a flare, as we do, provides several physical constraints by which many of the field lines may be discarded.  Of those that remain we find that the one minimizing the error metric, is the most plausible one, and therefore deem it the best fit.  The absence of such physical constraints when considering quiescent active regions forced \cite{Malanushenko2009} to define a particular local minimum as the best fit.

Of the 2000 field lines modeling a particular PRFL, we find that most do not satisfy basic properties of a flare loop, and therefore can be discarded.  First, we end the tracing if the field line crosses the upper boundary at $z=20$~Mm.  The maximum separation of the two flare ribbons is about 20 Mm (Figure~\ref{fig:overview}f), and we assume that no valid PRFL would exceed that height.  Next we require that a projected field line be longer, in the plane-of-the sky, than the observed loop in order to be valid. Since coronal loops are only partially visible in a particular EUV image (see Figure~\ref{fig:model}a) they are often shorter but never longer than the field line they follow.
Finally, we invoke the assumption that flare ribbons observed in AIA UV 1600~\AA\ constitute the feet of the PRFLs. 
We therefore require a valid field line to have both feet somewhere in the chromospheric flare ribbons. { Note that there are uncertainties in identifying flare ribbon pixels; in particular weakly brightened pixels might be missed \citep[see discussions in][]{Qiu2025}. Furthermore, the width of observed loops is typically a few AIA pixels. To take into account these factors, we smooth the flare ribbon mask (Figure~\ref{fig:overview}f) with a 4-pixel by 4-pixel boxing car, and require valid field lines to land inside the smoothed ribbon mask.}

These three criteria eliminate a significant fraction of the 2000 field lines from further consideration. { For a small number of loops, all 2000 field lines are eliminated, or there is not a valid fit to the loop. For example, a curvilinear structure lying on the northern ribbon in Figure~\ref{fig:overview}e (in green color) is indeed a section of the ribbon, or a false loop. These few false loops cannot pass the three criteria, and are therefore eliminated from the sample.} { The remaining field lines fitting a loop provide a sparse sampling of the $(\alpha,H)$ parameter space, from which the overall shape of the error function cannot be inferred. Therefore, out of the remaining field lines, we take the one minimizing the error metric to be the best fit, as described in the following text.}

\subsection{Error Metric}
\label{subsec:error}

We use an error metric, $\sigma$, to quantify the degree to which a magnetic field line, projected onto the plane-of-the-sky, accurately represents an observed PRFL.  Our error metric measures the average plane-of-the-sky separation between the field line and its target loop { \citep[also see][]{Warren2018}}, so the minimum value provides the best approximation.  

\begin{figure}
\begin{centering}
\includegraphics[width=.43\linewidth]{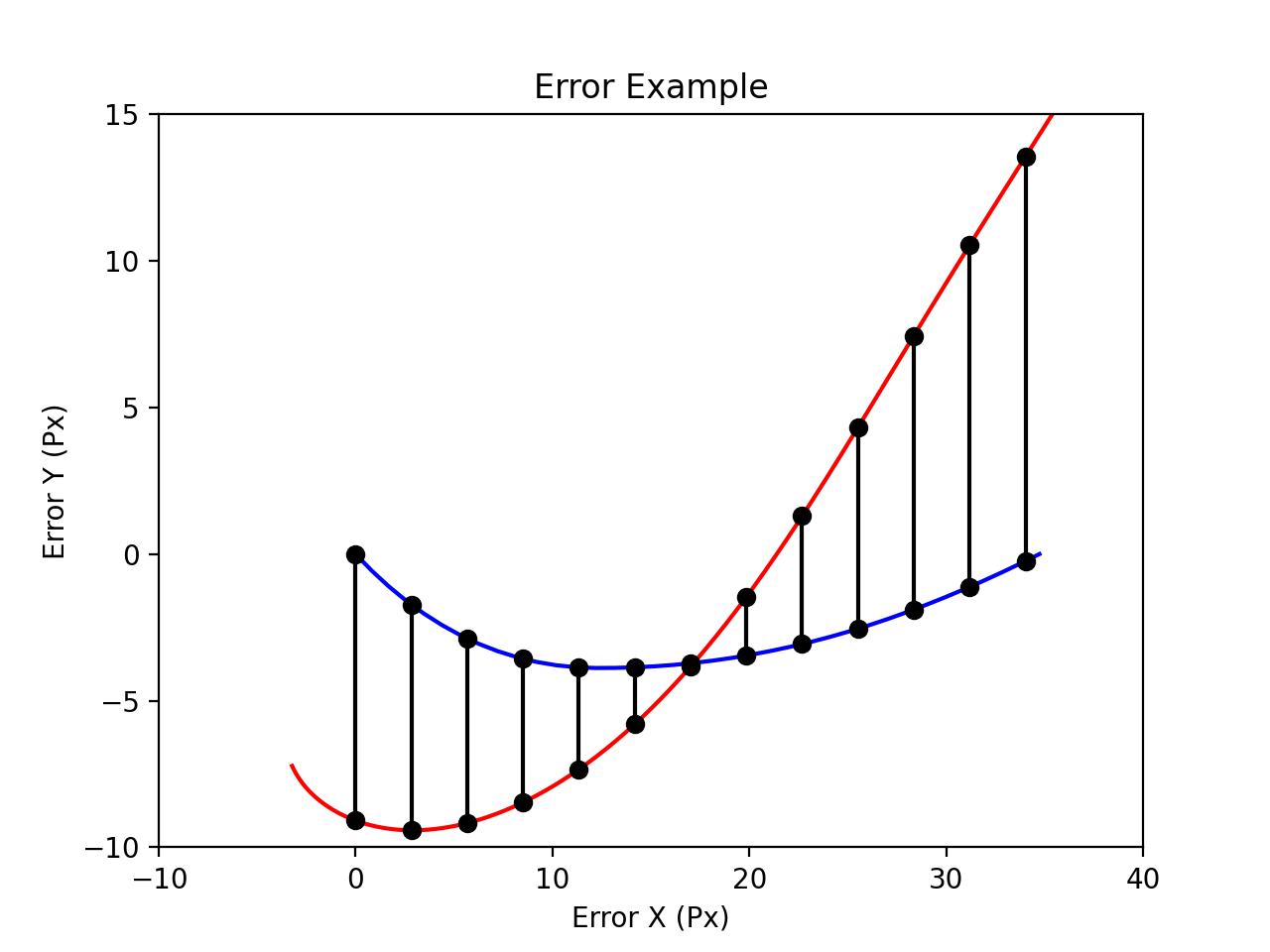}
\includegraphics[width=.46\linewidth]{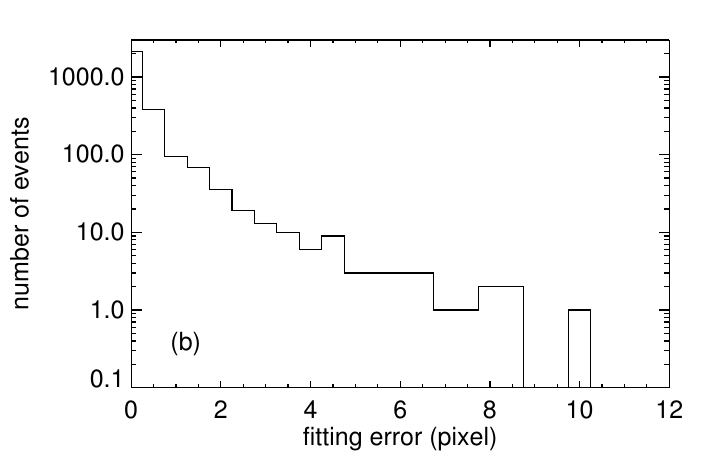}
\caption{(a) An example demonstrating the error metric between an observed loop (blue) and modeled field line (red).  The separation $\Delta Y(X)$ is shown by vertical black lines.
(b) Histogram of the fitting error $\sigma$ for 2807 loops.}
\label{fig:error_example}
\end{centering}
\end{figure}

We measure the separation as the distance between the two curves, along the direction ($Y$) perpendicular to the line connecting the ends of the observed PRFL.  This line is the $X$ axis in what we hereafter call ``error space'', and the loop's ends are located at $(0,0)$ and $(X_{\rm max},0)$, as illustrated by the blue curve in Figure~\ref{fig:error_example}.  Both curves are rotated into error space and fit with splines, as illustrated by the two curves in Figure~\ref{fig:error_example}a.  The separation distance, $\Delta Y(X)$, is computed at regularly spaced values of $X$ covering the full range of the PRFL: $0\le X\le X_{\rm max}$.  Because we have eliminated field lines shorter than the PRFL, the former will always extend beyond the latter in error space (compare red and blue curves in Figure~\ref{fig:error_example}a).  The error metric, $\sigma$, is defined as the average unsigned distance between the curves
\begin{equation}
  \sigma ~=~ {1\over X_{\rm max}}\int\limits_{0}^{X_{\rm max}} \Bigl| \Delta Y(X)\Bigr|\, dX ~~,
    \label{eq:sigma}
\end{equation}
approximated by the rectangle rule. 

Of all valid field lines for a given PRFL, we designate the best fit as that field line with the minimum value of $\sigma(\alpha,H)$.  Because we discarded most of the 2000 possible field lines, the $(\alpha,H)$ parameter space is sparsely sampled and cannot be searched using an algorithm assuming a smooth function.  On the other hand, there are few enough points that the minimum can be found through a full search.  In practice we minimize over $H$ for each value of $\alpha$ and then find the minimum of these minima.


\orcca\ fitting associates with every observed PRFL a 3D field line along with a value of twist, $\alpha$.  It only fails in cases (about 4\% in our study) where none of the 2000 possible field lines satisfy the criteria.
Figure~\ref{fig:error_example}b shows the distribution of the minimum $\sigma$ for the remaining 2807 loops (96\%), indicating that the mean separation between observed and model field line is well within 2-3 AIA pixels, which is the minimal observed loop width \citep{Aschwanden2017}. The very few loops with $\sigma > 3$ pixels are discarded. 

Of the remaining 2728 (94\%) loops that are successfully fit, 
each has a three-dimensional geometry and a value of $\alpha$.  As each one is generated from a different LFFF, they cannot be combined into an equilibrium field.  Nevertheless, we expect each field line to be a reasonable approximation, in three dimensions, of the actual PRFL after it has relaxed following reconnection \citep{Malanushenko2009}.  { In more than one half of the successful cases, the length of the projected best-fit field line is between 1 to 3 times the length of the observed loop. Since an observed loop is only part of a full loop, we subsequently develop a method to identify different parts of the same loop in \S3.3.}



\subsection{Identification of Unique PRFLs}
\label{subsec:consol}

Among the 2728 loops successfully fit by \orcca, there are undoubtedly instances of the same physical loop identified at multiple times.  
AIA observations at 12~s cadence are able to capture a single loop multiple times as its plasma cools through an EUV passband --- typically on a timescale of minutes. 
Sometimes, a loop may appear broken into multiple segments in a single image, which are recognized by OCCULT as several loops. Each observation of a given PRFL can therefore be fit multiple times by different modeled field lines. Identifying and consolidating cases where a single field line has been fit multiple times provides a better representation of the reconnection process. It also offers the opportunity of estimating the uncertainty in our new methodology. 

\begin{figure}
\begin{centering}
\includegraphics[width=.52\linewidth]{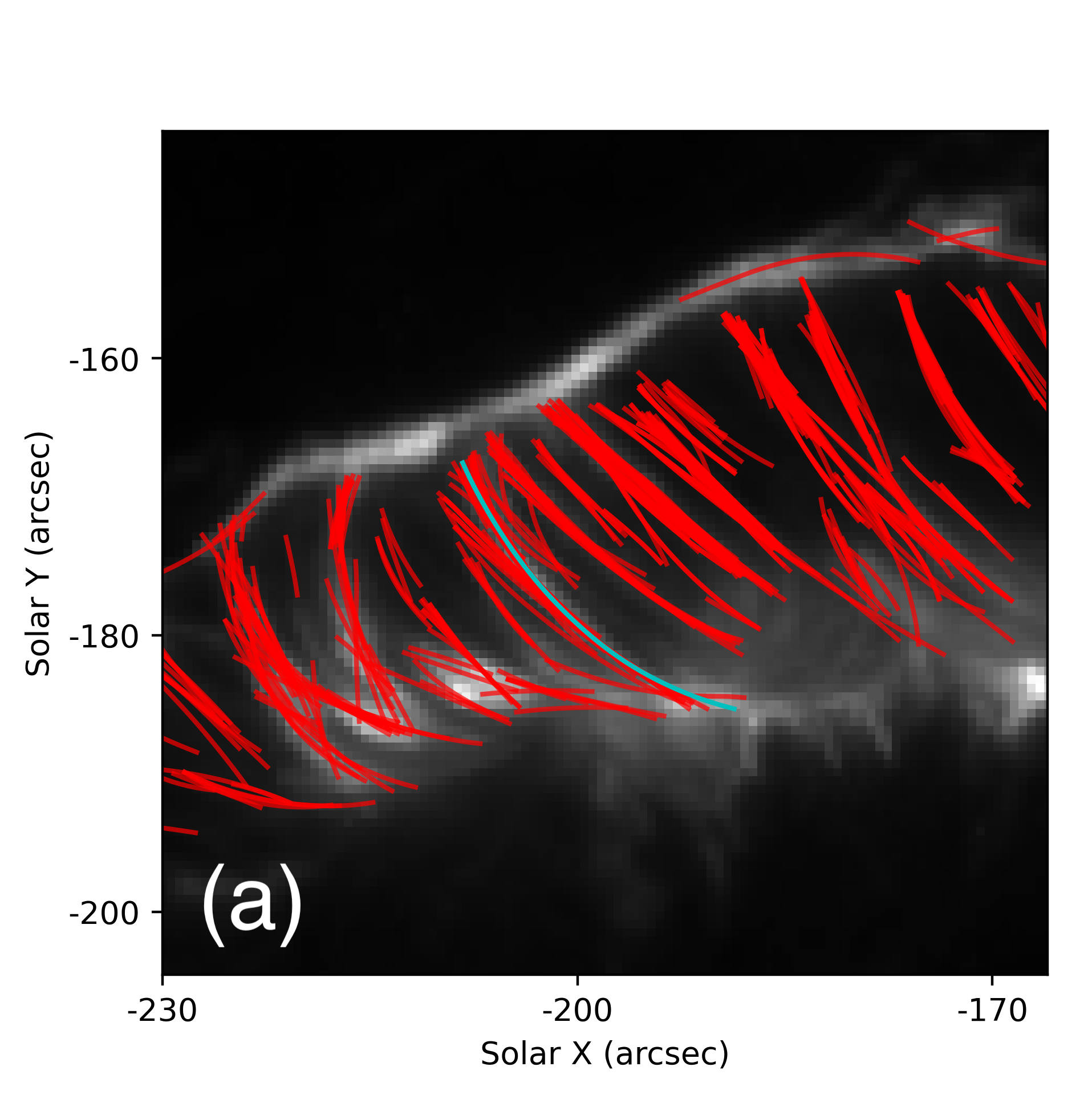}
\includegraphics[width=.46\linewidth]{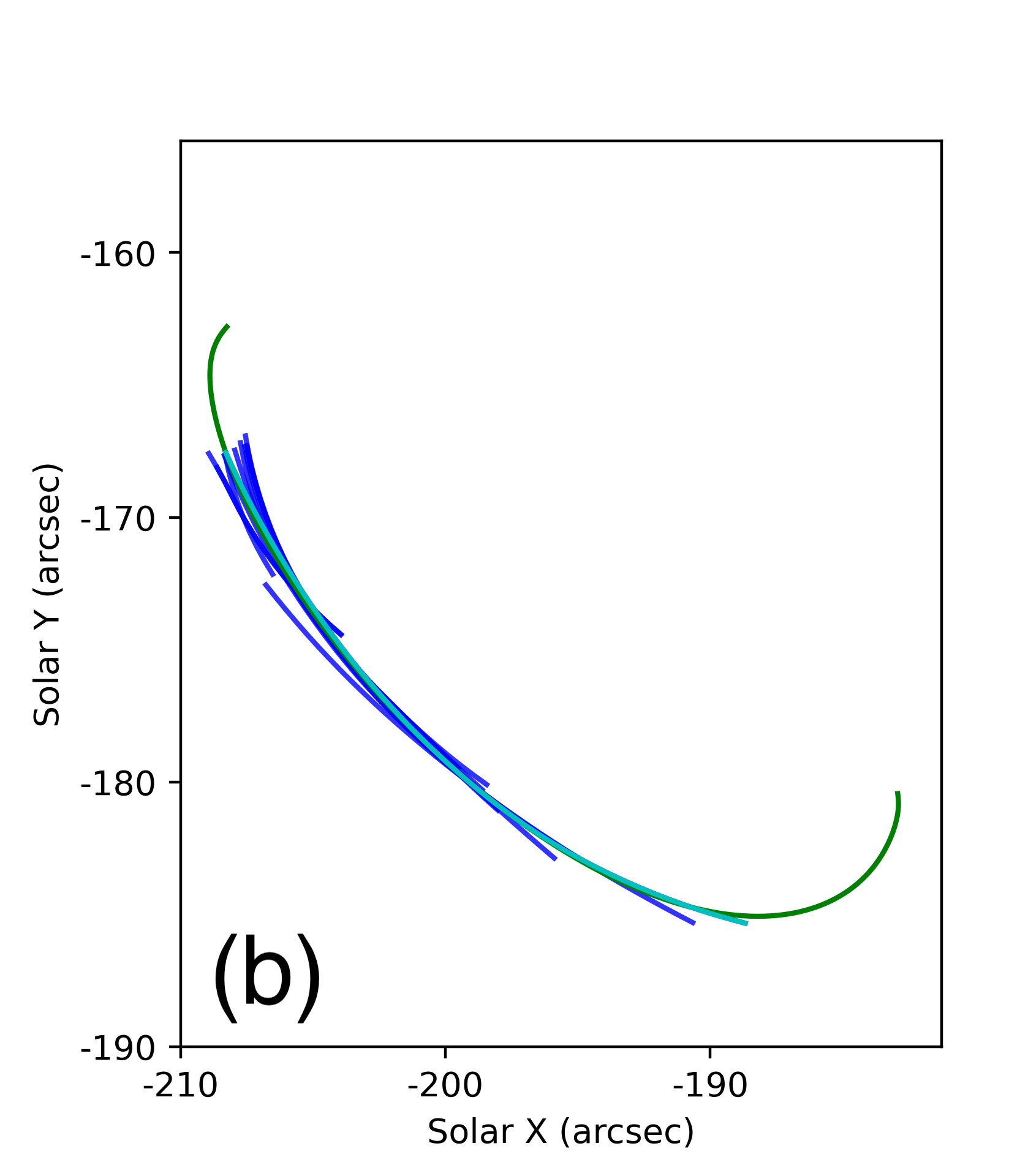}
\caption{(a) An example of several tens of observed loops in both AIA 304~\AA\ and 171~\AA\ passbands identified with the OCCULT code (red) within a time window of { four} minutes, superimposed on an AIA 304~\AA\ image. The loop in cyan is one of the reference loops within this time window. (b) A zoomed-in view of the consolidation, showing 12 observed loops (blue) that match a model loop (green) of the longest observed loop, or reference loop (cyan), with $\sigma \le \sigma_{thr}$.}
\label{fig:consolidation}
\end{centering}
\end{figure}

We attempt to identify and consolidate repeat observations of the same PRFL over a { 4-minute} window.  The time window is based on the expected time for a PRFL to cool through a particular passband \citep[e.g.][]{Aschwanden2001, Qiu2016}.  An example of all the observed loops in one time window is plotted in red in Figure~\ref{fig:consolidation}a, superimposed on an AIA 304~\AA\ image.  It appears possible that some of these are repeat observations of the same loop, or 
partial sections of that loop.

The model field line of the longest observed loop within the { 4-minute} window is taken as a reference, and compared to all others within the window to find any repeats.  It is compared using error metric, $\sigma_m$, defined in eq.\ (\ref{eq:sigma}), but computed between the reference field line and all other observed loops.  Each loop with $\sigma_m<\sigma_{\rm thr}$, is taken to be a repeat observation of the reference.  After removing from consideration the reference loop, and all repeat observations of it (if there are any), we repeat this procedure using the model field line of the longest of the {\em remaining} observed loops in that time window as the new reference loop.  This is iterated until no loops remain in the time window.

We have arrived at the optimal threshold, $\sigma_{\rm thr}=0.5$ pixels through experimentation.  We find this to match various loop sections while avoiding the previously formed, underlying PRFLs. This is a stringent condition considering the average observational width of the loops is greater than one AIA pixel \citep{Aschwanden2017}. With this choice of $\sigma_{\rm thr}$ we reduce the number of PRFLs from 2728, to 920 unique loops. 
In those cases where the reference field line fits its own loop with $\sigma<3$ pixel but $\sigma>\sigma_{\rm thr}$, a second match, i.e.\ repeat, is unlikely.  There are almost no repeats in such cases. 

Figure~\ref{fig:consolidation}b shows an example of this procedure at work. Among all the observed loops (red) in the time window, 12 of them (blue) are found to match the field line (green) of the longest observed loop (cyan), with $\sigma_m < \sigma_{\rm thr}$.  We take these to be 12 repeat observations of the same loop: reference loop, and hereafter consolidate them into a single field line. 

\begin{figure}
\begin{centering}
\includegraphics[width=.33\linewidth]{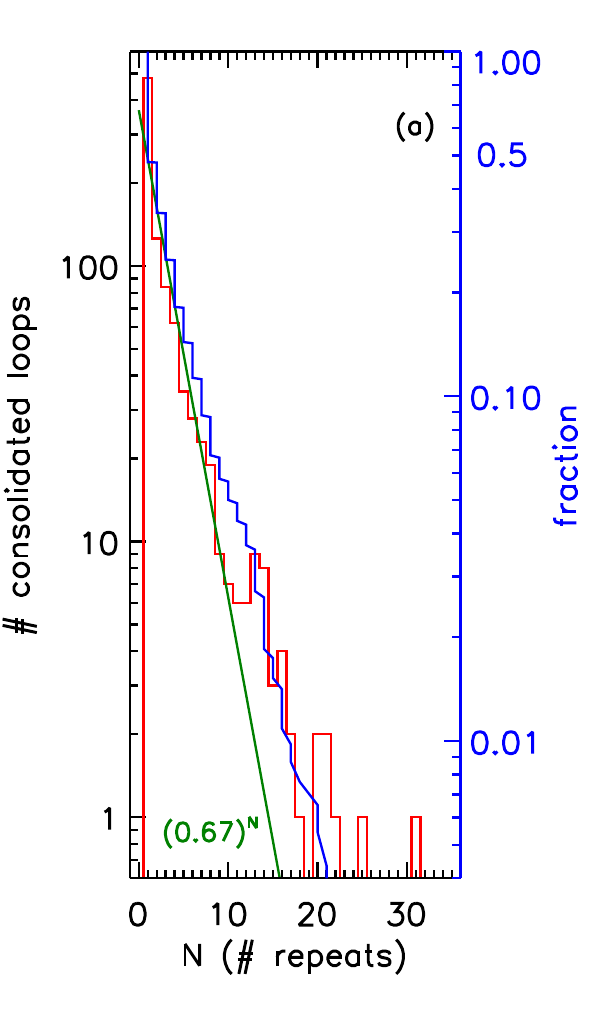}
\includegraphics[width=.66\linewidth]{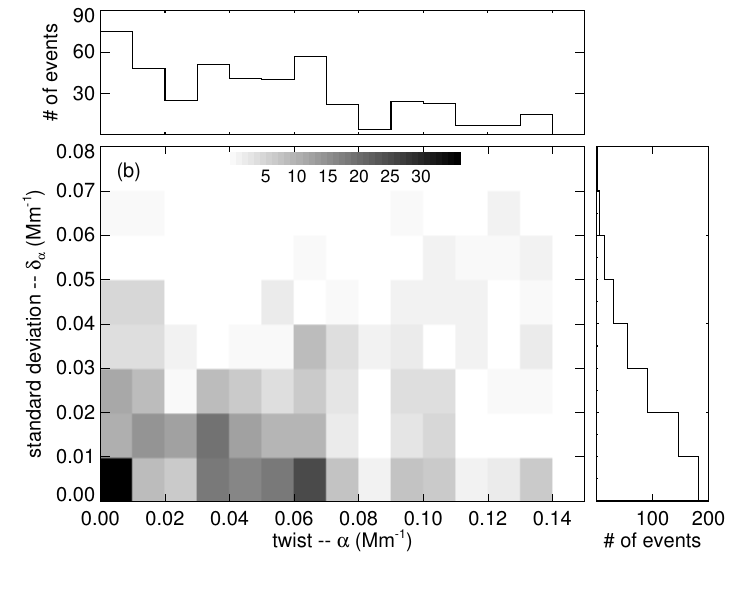}
\caption{(a) Histograms of repeat measurements. Red curve is the histogram of $N$, read from the left axis, with a green curve showing an exponential fit.  The blue curve, read from the right axis, is a cumulative histogram.  (b) Histograms of the twist $\alpha$ and its standard deviation $\delta_{\alpha}$ of the 439 loops that are observed for more than once ($N > 1$).}
\label{fig:repeat}
\end{centering}
\end{figure}

Figure~\ref{fig:repeat}a characterizes the consolidation process through a histogram of the number $N$ of repeat loops, including the reference itself (i.e.\ $N\ge1$). Among the 920 loops, 481 of them, or about one half, have been observed only once ($N=1$). Each of the other 439 loops appears more than once, as determined by the above algorithm. The histogram falls off geometrically as $\sim (0.67)^N$, which could occur if each successive image had a $67\%$ chance of satisfying the matching criterion.  The consolidation of 2728 loops into 920 groups gives an average group size of $\langle N\rangle\simeq3$, matching the mean of the normalized geometric series. 


For each of the 439 loops with repeat observations ($N>1$) the fitting provides multiple measurements of the same quantity.  We adopt the value from the reference loop for the consolidated loop, and use the other values to estimate measurement uncertainty.  For example, the standard deviation  in $\alpha$, denoted as $\delta_{\alpha}$,  provides an estimate of measurement uncertainty in $\alpha$. Figure~\ref{fig:repeat}b is a histogram of the 439 loops over $\alpha$ -- $\delta_{\alpha}$ space.  It shows that $\delta_{\alpha}<0.02\,{\rm Mm}^{-1}$ for more than 60\% of the loops and $\delta_{\alpha} <0.04\,{\rm Mm}^{-1}$ for more than 90\% of the loops.  The standard deviation $\delta_{\alpha}$ is not strongly correlated with $\alpha$ itself.

\begin{figure}
\begin{centering}
\includegraphics[width=.39\linewidth]{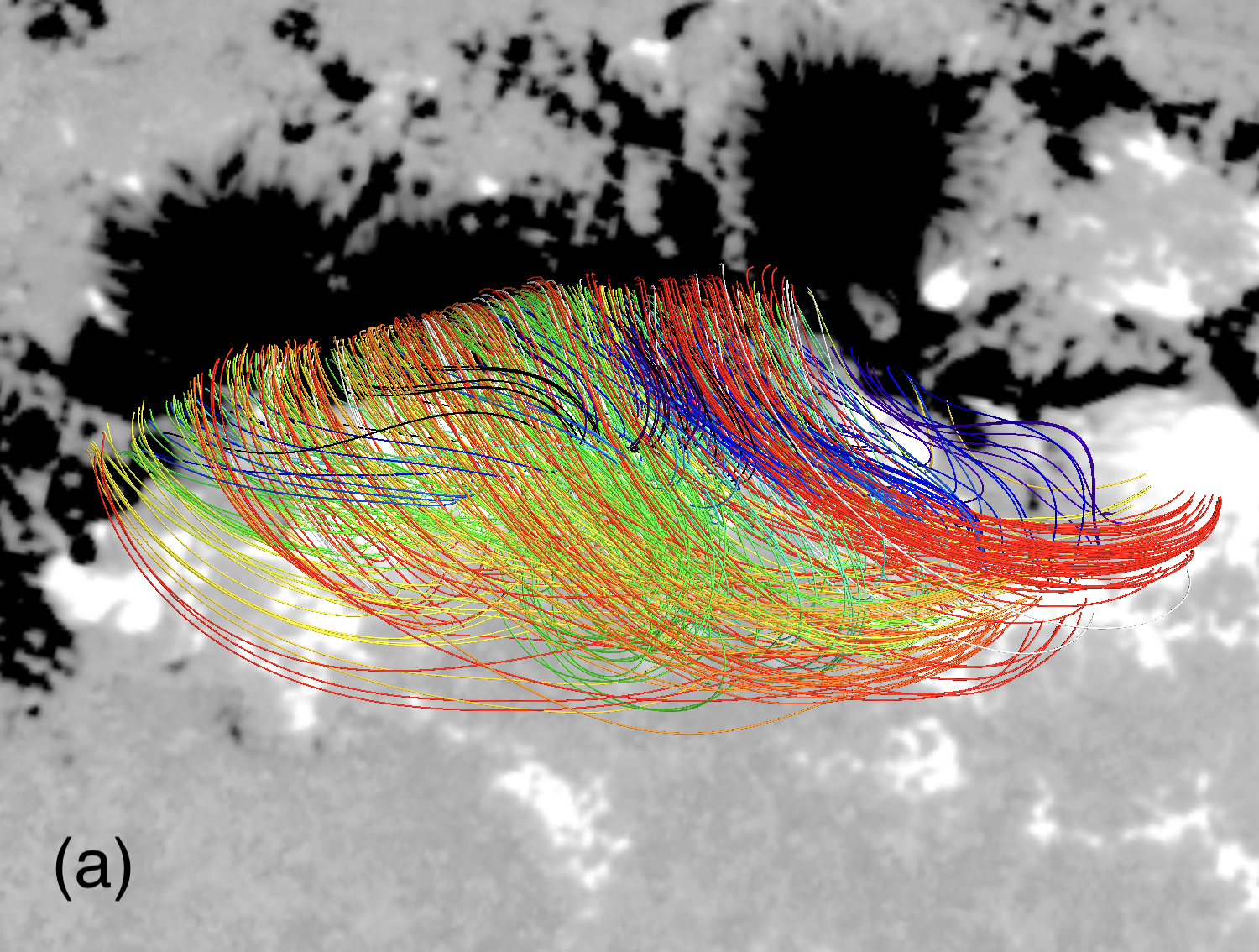}
\includegraphics[width=.44\linewidth]{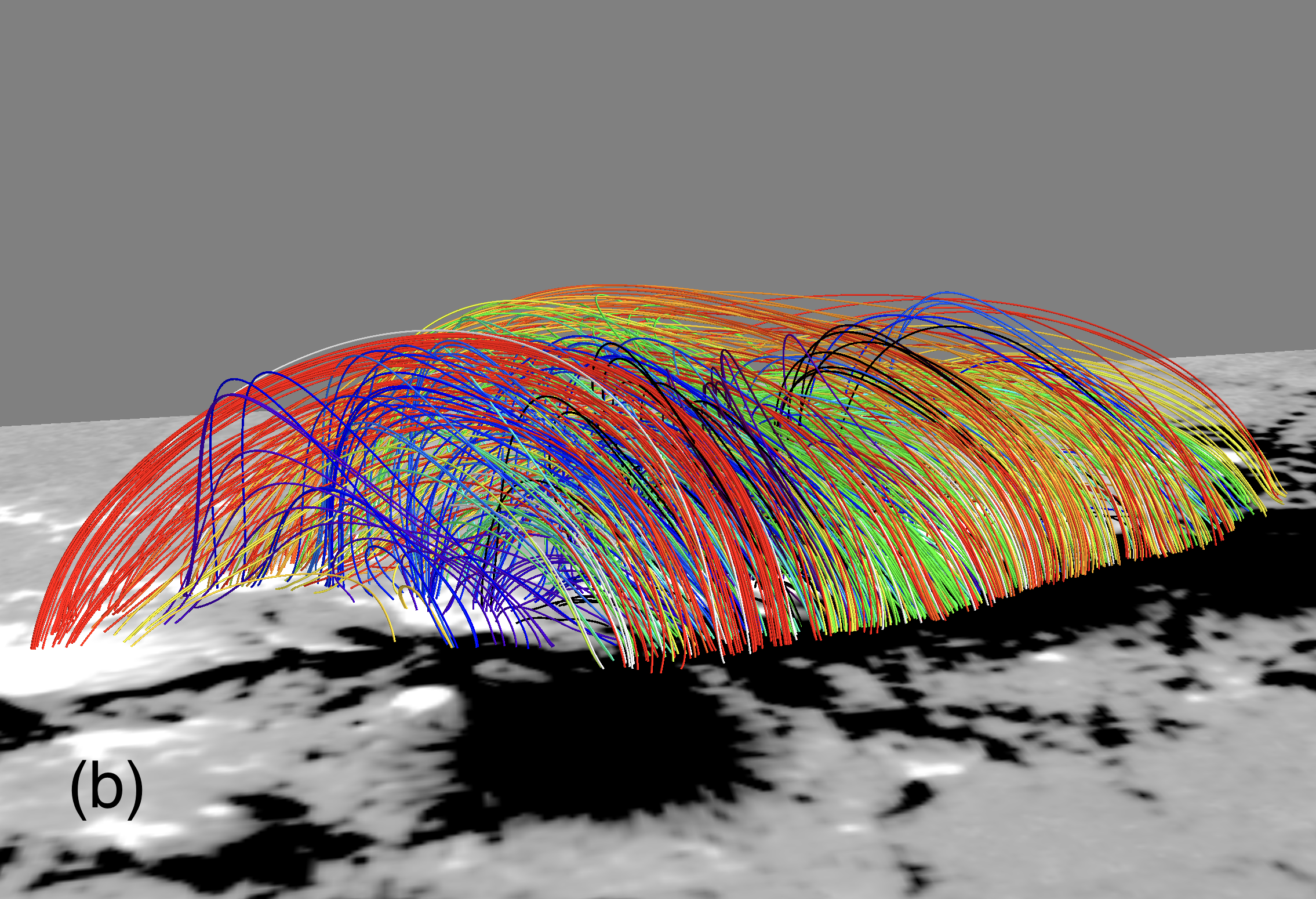}
\includegraphics[width=.055\linewidth]{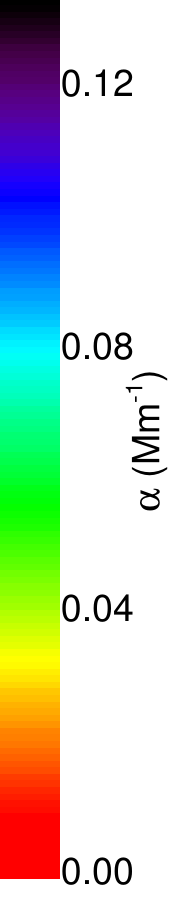}
\caption{Top (a) and side (b) view of the model field lines of the unique reference loops. Colors indicate their $\alpha$ values illustrated in the color bar.}
\label{fig:allloop}
\end{centering}
\end{figure}

Finally, Figure~\ref{fig:allloop} shows the field lines of all unique loops, consolidated from nearly 3,000 observed loops, viewed from the top (a), and from the side (b). Here the colors indicate the $\alpha$ values from the \orcca\ fit. It is shown that loops anchored in the photosphere closer to the PIL tend to have larger $\alpha$ values. These properties will be examined in detail in the next section.  

{ The \orcca\ method provides this 3D information for each field line.  It does not necessarily produce a self-consistent 3D field over the entire volume.  As a consequence, some field lines will cross one another, even though this cannot occur in a genuine, force-free field.  This appears to affect a small number of field lines shown in Figure ~\ref{fig:allloop}.}   


\section{Properties and Evolution of PRFLs}
\label{sec:results}

\begin{figure}
\begin{centering}
 \centering
    \includegraphics[width=0.48\textwidth]{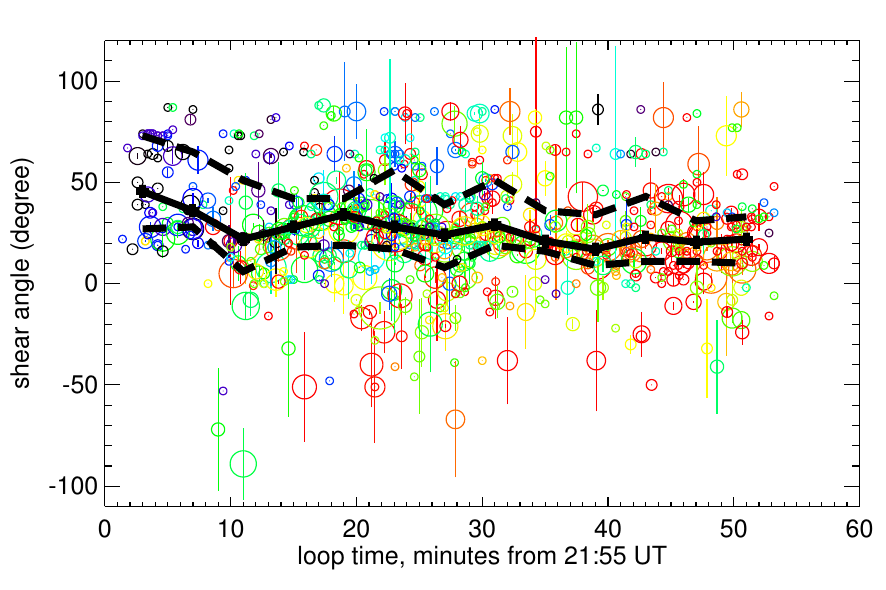}%
    \includegraphics[width=0.48\textwidth]{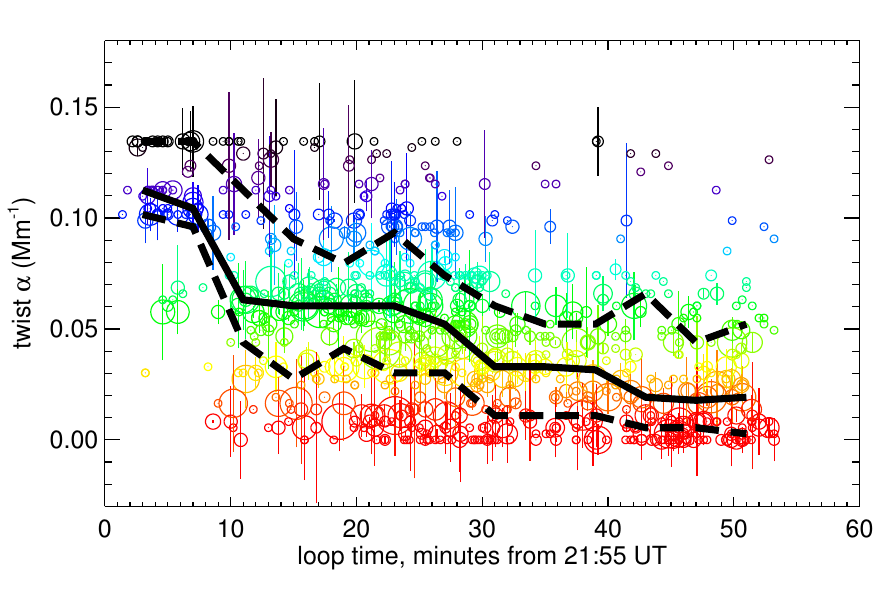}%
    \\
    \parbox[c]{3.3in}{    (a)} \parbox[c]{3.3in}{    (b)} \\
    \includegraphics[width=0.48\textwidth]{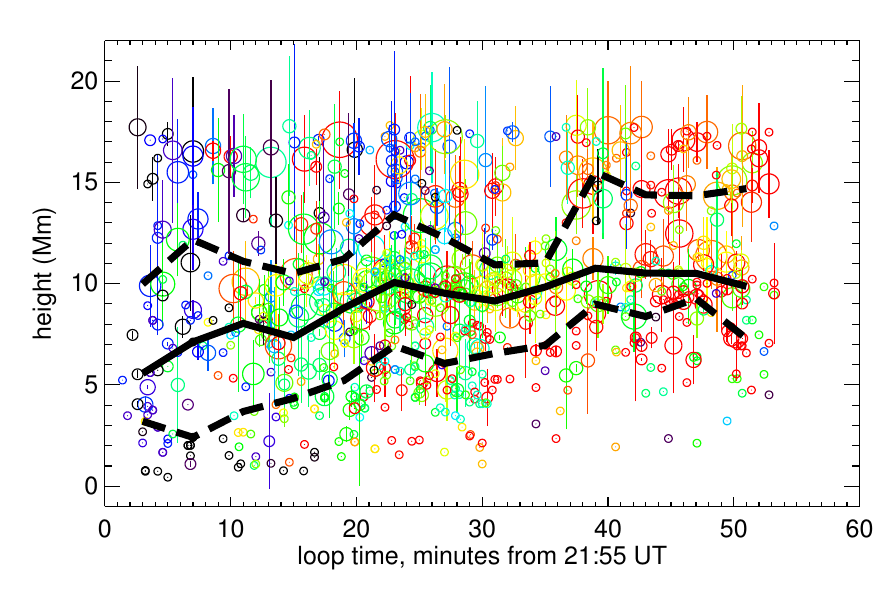}%
    \includegraphics[width=0.48\textwidth]{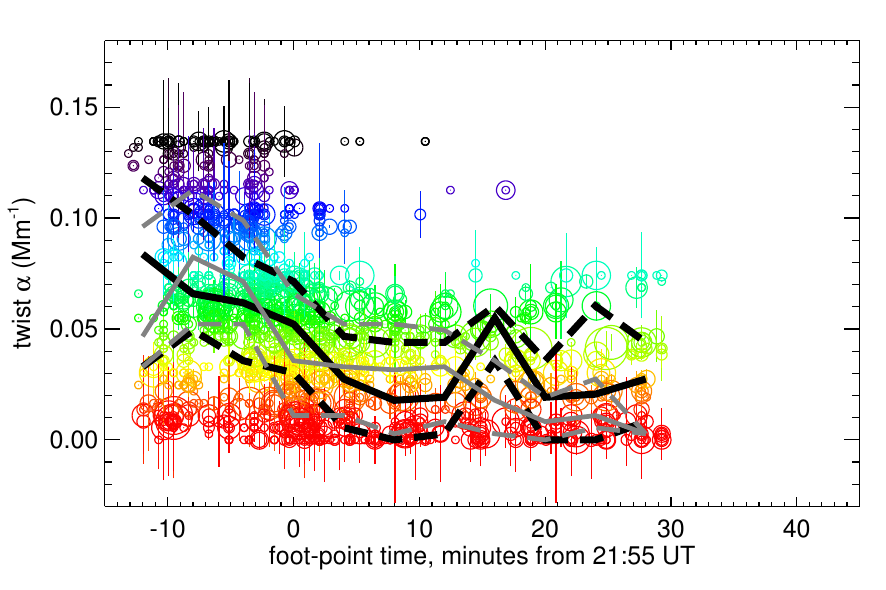}%
    \\
    \parbox[t]{3.3in}{    (c)} \parbox[c]{3.3in}{    (d)} 
\caption{Properties of PRFLs, including the shear angle with the PIL (a), twist parameter $\alpha$ (b), and height $z_{max}$ (c) of unique model loops against the time when loops are observed. In each panel, the data points are plotted in color circles, with the color indicating the twist value, given in the color bar in Figure~\ref{fig:allloop}. The area of the circle is proportional to the number of repeat observations for each unique loop, and the vertical bar of the same color indicates one half of the standard deviation of the measurements of the repeat loops. The thick solid { black} line shows the median value within a { 4-minute} window and the thick dashed { black} lines show the quartiles of the values in the window. (d) Twist parameter $\alpha$ against the time of the foot-point brightening (see \S\ref{sec:overview} for definition) in both positive and negative magnetic fields. The median and quartiles of $\alpha$ values within every { 4 minutes} are plotted in thick { black}/{ gray} lines for the positive/negative foot-point timing. }
\label{fig:results}
\end{centering}
\end{figure}

We have obtained the 3D quasi-force-free magnetic field describing the PRFLs, with which we characterize the physical nature of flare evolution, including the strong-to-weak shear evolution. Figure \ref{fig:results} shows properties of PRFLs during the entire course of the flare, derived for the 920 consolidated loops, i.e.\ unique field lines. The properties, including the shear (Figure~\ref{fig:results}a), twist (Figure~\ref{fig:results}b, d), and height (Figure~\ref{fig:results}c), of each unique model loop are plotted as open circles, with the color indicating the $\alpha$ value from large (violet) to small (red). The area of each circle is proportional to the number of repeat observations ($N$ from 1 to 31), and the vertical bar denotes the standard deviation of the measurements (see \S\ref{subsec:consol}). Finally, the median values of each property, measured from the 920 unique field lines, over a { 4-min} window is plotted as a thick line, with corresponding dashed curves showing the quartiles over the window\footnote{The median values of the measurements from all 2728 loops are not significantly different, and are therefore not displayed here.}.


The conventional characterization of shear angle, $\theta_{lp}$, measures the angle of the observed loop with respect to the PIL of the photospheric radial field.  Figure \ref{fig:results}a shows this value, but using the best-fit model field line projected onto the image plane.  { The 2D shear angle is measured from the normal vector $\hat{\bf n}$ of the local PIL, pointing from positive to negative polarities, to the local field line vector $\hat{\bf f}$ also pointing from positive to negative polarity. Positive angles show the measurement from $\hat{\bf n}$ to $\hat{\bf f}$ counterclockwise, and negative angles show the measurement clockwise.  We note that $\theta_{lp}$ is measured with respect to the PIL of the photospheric magnetogram, which often has a complex shape rather than a straight line. The curvature of the PIL introduces ambiguities in $\theta_{lp}$ measurements, leading to, for example, some of the very large shear angle $|\theta_{lp}| \approx 90^{\circ}$ in Figure~\ref{fig:results}a. Despite the scattering due to the measurement ambiguity, overall, the shear angle $\theta_{lp}$ is dominantly positive, reflecting a globally right-handed shear.} Furthermore, it exhibits a decrease over time consistent with strong-to-weak shear evolution reported by \citet{Qiu2023}.  In the present case the median $\theta_{lp}$ ({ black} curve) decreases from more than 40 degrees to about 20 degrees over 50 minutes.

The shear angle $\theta_{lp}^p$ of {\em potential field lines} traced from flare ribbons have been measured in the previous study \citep{Qiu2023}. In a realistic configuration, the shear angle of the potential field is not zero, and may fluctuate by 10 to 20 degrees \citep[also see][]{Qiu2017}. The upshot is that the observed strong-to-weak shear evolution of PRFLs {\em does not} necessarily indicate an approach to a potential field, or relaxation. 

The \orcca\ fitting technique provides a more physically grounded measurement of magnetic shear in the form of the twist parameter $\alpha$.  This quantity explicitly measures the field's departure from a potential field, since $\alpha=0$ is potential.  The evolution of $\alpha$, shown in both Figures \ref{fig:results}b and \ref{fig:results}d, exhibits strong-to-weak evolution as $\alpha$ decreases over 50 minutes from 0.12 to about 0.02 Mm$^{-1}$.  This makes a 
compelling case that later formed PRFLs are closer to a potential configuration.  Moreover, it suggests a physical explanation for the strong-to-weak evolution in this two-ribbon flare: earlier formed loops have a larger twist, or a larger current per unit magnetic flux, than later formed loops. This is the first time that the physical nature of the long observed strong-to-weak shear evolution has been explained using observation-constrained 3D model.

{ It is notable that the traditional shear angle, $\theta_{lp}$, turns out to correlate, to some extent, with $\alpha$.  The Spearman correlation between $\theta_{lp}$ and $\alpha$ is $\rho=0.48$, among the consolidated loops.  While this is not strong, it has virtually no probability of arising by chance.  Thus, the empirical measure of shear, though with measurement ambiguities, is found to have some relation to the actual shear of the post-reconnection magnetic field.}

Flare loops are observed in EUV passbands after they have been filled and heated by chromosphere evaporation and then cooled down to a few million K to be visible in these images. This process leads to a significant time lag between the observed signatures of reconnection and heating, i.e.\ the foot-point brightening, and loop's appearance in EUV.   The delay makes it very difficult to identify the feet of PRFLs. Now in possession of a 3D model of the PRFLs, we can track each loop to its conjugate feet in the lower atmosphere, and identify the time of its formation by reconnection.

\begin{figure}    
    \centering
    \includegraphics[width=0.8\textwidth]{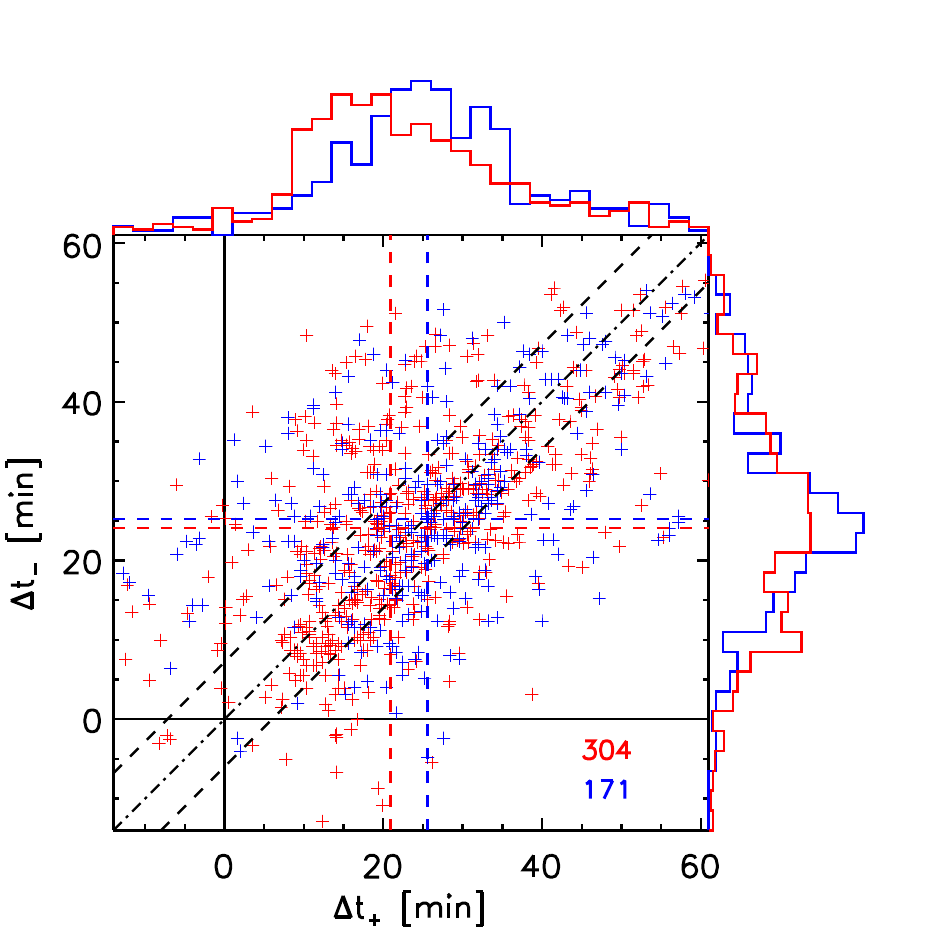}%
    \caption{Plot of time delays $\Delta t_+$ and $\Delta t_-$.  The central scatter plot shows $\Delta t_-$ and $\Delta t_+$ for each reference loop as a cross: red for 304\AA, and blue for 171\AA.  Vertical and horizontal dashed lines of matching color show the medians. A broken black line indicates $\Delta t_+=\Delta t_-$, and dashed diagonals above and below it mark off the central quartiles of the distribution.  Normalized histograms of each delay are plotted along the side and top in matching color for each wavelength.}
    \label{fig:delay}
\end{figure}

Using our 3D field lines we associate each coronal loop observed in a particular EUV band with the chromospheric footpoints which brighten in AIA 1600\AA\ to form the flare ribbon.  The difference between the time of footpoint brightening and loop appearance defines a time delay $\Delta t$.  Indeed, since each EUV loop is paired with both a positive and a negative footpoint, we measure two delays, $\Delta t_+$ and $\Delta t_-$, which should theoretically be the same.  Figure \ref{fig:delay} shows the distribution of these two measurements for each of the 920 reference loops, 570 fit from 304\AA\ (red) and 350 fit from 171 \AA\ (blue).  The scatter plot shows that the two delays, $\Delta t_+$ and $\Delta t_-$, are strongly correlated (Spearman $\rho=0.56$) though do not agree precisely. Most delays differ by less than 7 minutes (see dashed diagonals), which we take as the measurement uncertainty.  
Furthermore, we see that both wavelengths exhibit very similar distributions, with delays of roughly 24 minutes between reconnection (i.e.\ footpoint brightening) and loop appearance in the EUV passbands characterizing plasma temperature $\sim$ 1MK. 
The implication of these results will be further discussed in \S\ref{subsec:cooling}.

Figure \ref{fig:results}d plots the twist of consolidated loops {\em vs.} the brightening time of their associated footpoint, in the same format as Figure \ref{fig:results}b.  Since each loop has two footpoints, it is represented by two circles on the plot. This also leads to two windowed-median curves, { black} and { grey}.  Comparison between the figures shows that strong-to-weak evolution appears better organized by time of reconnection (d) than by time of loop appearance (b).  The time axis is also displaced leftward on (d), beginning at negative time, due to the time delay.

\begin{figure}    
    \centering
    \includegraphics[width=0.98\textwidth]{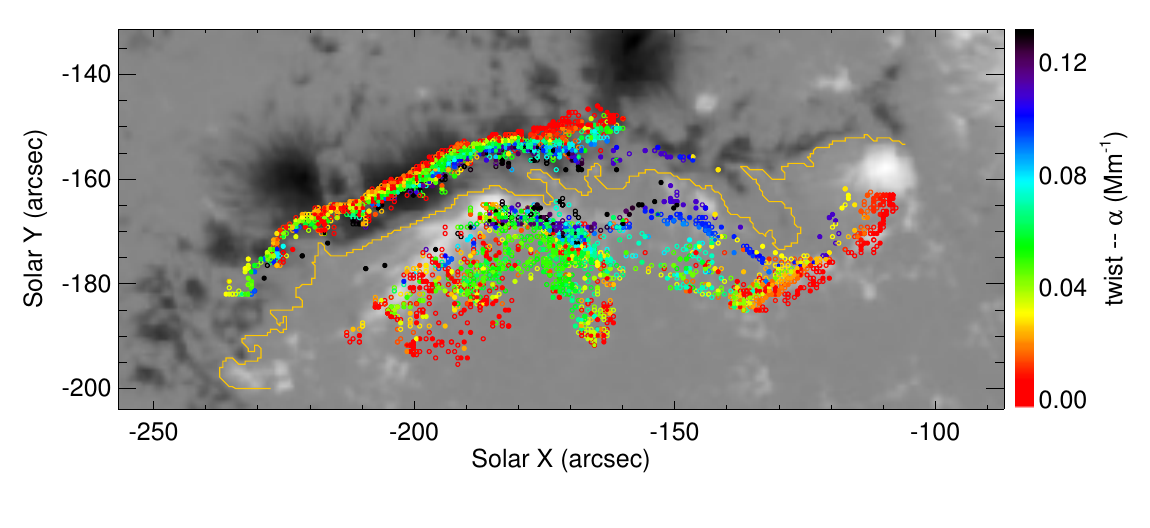}%
    \caption{Distribution of $\alpha$ at $z = 0$ for all best-fit model loops (open circle) and for the consolidated loops (filled circle), superimposed on a magnetogram of the radial component of the photospheric magnetic field in helioprojective coordinates (AIA image plane). The orange curve outlines the PIL of the radial magnetic field.}%
    \label{fig:feet}%
\end{figure}

The 3D model also reveals the outward ribbon motion coincides with decreasing in PRFL twist. Figure~\ref{fig:feet} displays values of $\alpha$ at the feet of the full 3D loops over a magnetogram (grey scale).  The spatial distribution exhibits a rather organized pattern with larger $\alpha$ (violet to blue) lined along the inner edges of the ribbons close to the PIL.  As the feet ``move" away from the PIL, and from each other, $\alpha$ becomes smaller (green to red) . The spatial distribution of $\alpha$ inside flare ribbons follows a pattern similar to that of flare ribbons spreading out over time (Figure~\ref{fig:overview}f), further supporting that later formed loops are taller with a larger foot-point separation, and are closer to the potential field.

In the standard picture of flare evolution, i.e.\ the CSHKP model, ongoing reconnection produces loops of increasing height.  The association of 3D properties with flare loops provided by \orcca, allows us to quantitatively corroborate this picture as in Figure \ref{fig:results}c showing the maximum height $z_{max}$, measured from the surface. The height of the loops increases over time, consistent with the standard picture. This also matches off-limb observations of flare arcades that grow in height \citep[e.g.][]{Masuda1994, Forbes1996, Longcope2018}. 




In summary, we have used our new \orcca\ fitting technique to show that the magnetic field following a flare is not potential nor is it linear force free. Instead the magnetic field is best described by a quasi-nonlinear force free field with earlier formed loops having a larger force-free factor $\alpha$ than later formed loops. This also provides, for the first time, a physical explanation for the strong-to-weak shear evolution in this two-ribbon flare.

\section{Discussions and Conclusions}\label{sec:summary}
\label{sec:conclusion}

\subsection{Summary of Findings}
\label{subsec:summary}

The strong-to-weak shear evolution of post reconnection flare loops (PRFLs) have been reported in observations for two decades. Without reliable three-dimensional models of post-reconnection magnetic field, the physical nature of the observed strong-to-weak shear evolution has not been clear.

In this study, we have developed a novel method to construct 3D magnetic field of PRFLs. A large number of PRFLs identified in EUV images are fitted by magnetic field lines, with the constraint that these field lines must be anchored in flare ribbons observed in UV images. This Optimized Ribbon-Constrained Constant-$\alpha$ (\orcca) fitting method is successfully applied to an eruptive two-ribbon flare which exhibits strong-to-weak shear evolution. 

With the 3D model, we find that PRFLs formed earlier have larger twist $\alpha$ than later formed PRLFs. During the flare evolution, as the two ribbons outlining the conjugate feet of the PRFLs expand away from the PIL and from each other, the PRFLs are taller, and their twist $\alpha$ decreases. This study therefore provides, for the first time, a physical explanation of the long-observed strong-to-weak shear evolution using the observation-constrained model.

With the \orcca\ capability to trace field lines of PRFLs to their feet in the chromosphere, we also find the time delay between foot-point brightening and appearance of PRFLs in the EUV passbands. In this flare, the median time delay of hundreds of PRFLs emitting at $\sim$1~MK is about 20~min.  This is consistent with previous measurements of cooling delays, but our use of \orcca\ fitting provides the first case where the delay is measured for individual loops in a flare.

These results, and the \orcca\ technique enabling these findings, have important implications on post-reconnection magnetic modeling and along-the-loop hydrodynamic modeling. The progress in these modeling efforts will improve our understanding of flare energetics.

\subsection{Implications on Post-Reconnection Magnetic Modeling}
\label{subsec:twist}

Prior to this study, vector magnetograms have been analyzed to derive parameters characterizing the shear or twist in an active region. \citet{Kazachenko2022} have analyzed vector magnetograms of the active region hosting the flare in this study, and have found strong vertical electric currents $I_z$ and shear $\mathcal{S}_{ph}$ adjacent to the polarity inversion line separating the two flare ribbons. Note that shear by is defined \citet{Kazachenko2022} as 
\begin{equation}
\mathcal{S}_{ph} \equiv |B| \cos^{-1}\left(\frac{{\bf B}\cdot{\bf B}^p}{BB^p}\right) {\rm (G\ degree)},
\end{equation}
a parameter measuring the difference between the local vector magnetic field $\bf{B}$ and the potential field ${\bf B}^p$ extrapolated from the same vertical component $B_z$. The analysis was conducted using magnetograms at 21:41 UT, on the eve of the flare occurrence.

\begin{figure}
\begin{centering}
\includegraphics[width=.98\linewidth]{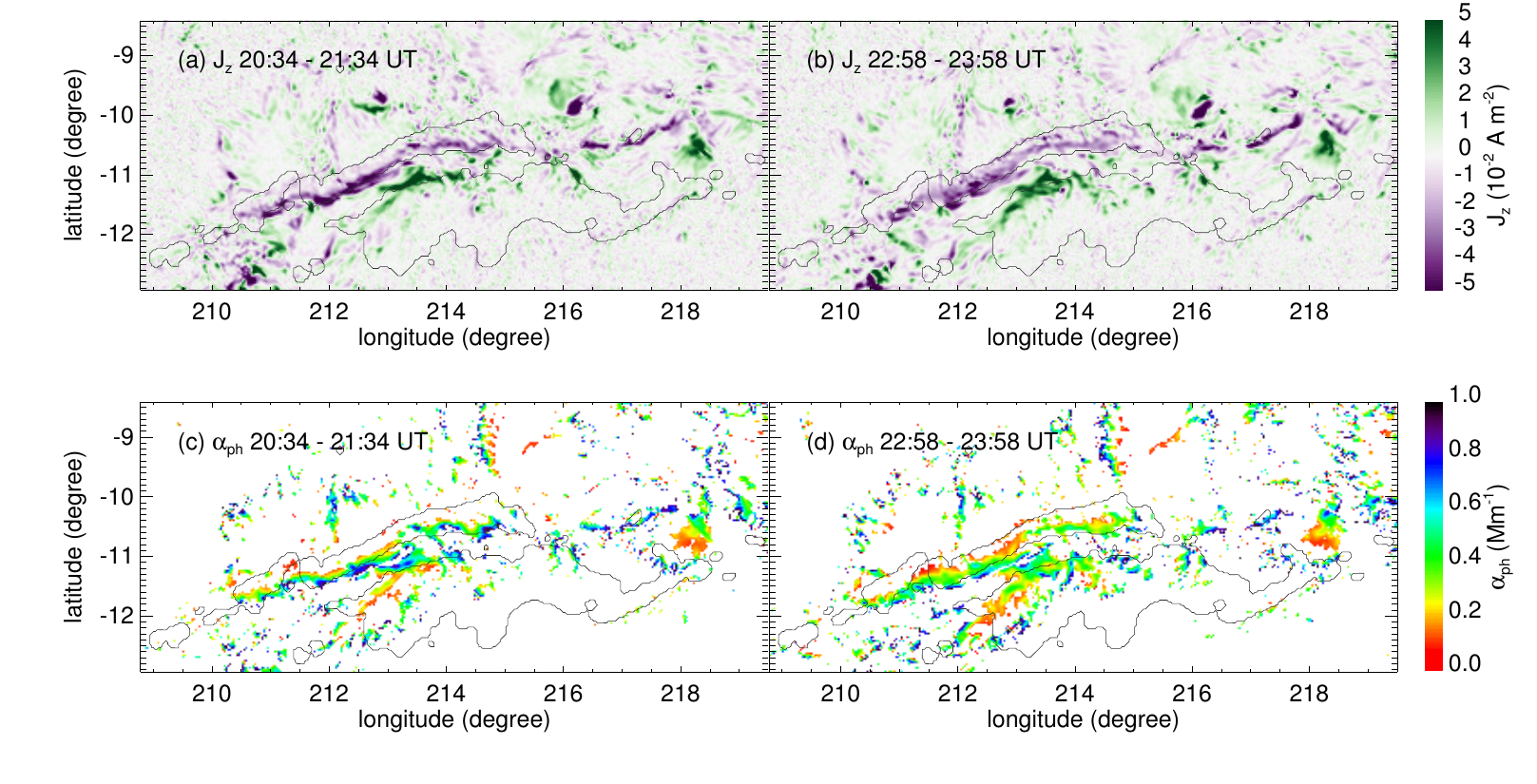}
\caption{Top: distribution of vertical electric currents averaged for 72 min before (a) and after (b) the flare, respectively. Bottom: distribution of $\langle \alpha\rangle$ averaged for 72 min before (c) and after (d) the flare. The black contours denote the flare ribbons, same as in Figure~\ref{fig:overview}f. The maps are casted in CEA coordinates of heliographic longtitudes and latitudes.}
\label{fig:hmi}
\end{centering}
\end{figure}

To place the $\alpha$ of PRFLs from our study in the context of the active region evolution, we derive the vertical electric current and force-free parameter $\alpha_{ph}$ from vector magnetograms of the photosphere, from three hours before the flare to three hours after the flare. During this period, the variation of the vertical magnetic field or flux is negligible (by 3\% within the active region, and 6\% within flare ribbons). The vertical currents and $\alpha_{ph}$, on the other hand, exhibit non-negligible changes after the flare.
Figure~\ref{fig:hmi} shows the distribution of the vertical electric current $j_z$ and the force-free parameter $\langle \alpha_{ph} \rangle$ before and after the flare. Here the computation of $\langle \alpha_{ph} \rangle$ is weighted by the vertical magnetic field $B_z$ and averaged within 3 by 3 pixels to minimize large errors associated with weak $B_z$, 
\begin{equation}
\langle \alpha_{ph} \rangle = \frac{\int (\nabla\times{\bf B})_z B_z ds}{\int B_z^2ds}.
\end{equation}

It is shown that strong vertical currents and large twist are distributed along the PIL. After the flare, the total currents have grown from 2.6 to 3.2 $\times 10^{12}$ A in the positive ribbon, and from -3.4 to -3.9 $\times 10^{12}$ A in the negative ribbon. Nevertheless, the peak current density has decreased after the flare, and currents appear to spread away from the PIL, and become more diffused into the ribbons. Similar pattern has been observed in the $\langle \alpha_{ph} \rangle$ distribution, with $\langle \alpha_{ph} \rangle$ becoming smaller yet more diffused inside the flare ribbons after the flare. 
The evolution of the active region twist or currents reflect re-organization of magnetic fields by flare reconnection. On the other hand, the $\langle \alpha_{ph} \rangle$ values are much larger than the PRFL-$\alpha$ determined with the \orcca\ method. The discrepancy may be physical: $\langle \alpha_{ph} \rangle$ characterizes non-potentiality in the photosphere, whereas PRFL-$\alpha$ describes the magnetic field in the corona.

{ \citet{Liu2023} have recently examined evolution of 21 active regions producing X-class flares (i.e., not including the active region in this study). They have measured $\langle \alpha_{ph} \rangle$ in a flare-mask determined from Q-maps, and reported that, after the flare, $\langle \alpha_{ph} \rangle $ appears to be more smoothly distributed within the flare mask, similar to the finding in this study. However, \citet{Liu2023} studied different active regions, determined the flare-mask with a method different from ours, and computed $\langle \alpha_{ph} \rangle$ with a smoothing box of different size; therefore, the $\langle \alpha_{ph} \rangle$ values presented in \citet{Liu2023} should not be directly compared with the values in this paper.}

In summary, in this flare, the PRFL arcade is anchored in a region with strong electric currents and twist as measured in the vector magnetograms. The comparison suggests that, qualitatively, the PRFL $\alpha$ distribution reflects the shear distribution originally built in the pre-reconnection field, perhaps in the form of a sheared-arcade \citep{Joshi2017, Prasad2023, Dahlin2022}. Measured by either $\alpha_{ph}$ or PRFL-$\alpha$, flare reconnection does not completely relax the field to a linear force free field \citep{Antiochos2002, Nandy2003, Longcope2008, Sun2012, Liu2023}. The \orcca\ method developed in this study takes advantage of observations in the corona, chromosphere, and photosphere, and models the coronal magnetic field with the {\em connectivity} established from observations of flare ribbons and loops. Therefore, the $\alpha$ distribution derived with this method provides crucial information that can lead to improved post-reconnection magnetic models and henceforth a better estimate of magnetic energy released in flares. This will be pursued in our future work.

\subsection{Implications on Hydrodynamic Modeling}
\label{subsec:cooling}
A { significant amount} of free magnetic energy released by flare reconnection is used to heat flare plasmas, producing prominent X-ray and EUV emissions in PRFLs and brightening at their feet. To date, we do not fully understand how plasmas in flare loops are heated to more than 10~MK and then cool down to 1~MK. One-dimensional along-the-field hydrodynamic simulations have been used to model this process \citep[e.g.][]{Warren2006, Longcope2014, Allred2015, Kowalski2017, Reep2020, Kerr2020}. In solar flares, heating and cooling of the chromosphere and corona have to be studied coherently. Therefore, observational signatures from the chromosphere (flare ribbons) to the corona (PRFLs) along the {\em same} loop will provide crucial tests on these models. However, it has been challenging to identify the {\em same} loop from its conjugate feet in the chromosphere to the corona. The \orcca\ method makes a significant progress to overcome this long-standing obstacle. It traces an entire flare loop from the chromosphere to the corona, along which, the tempo-spatial evolution of plasma emission in multiple wavelengths can be measured to provide observational test on hydrodynamic models. As an example, Figure~\ref{fig:delay} shows the time delay of EUV emission in PRFLs with respect to their foot-point brightening, suggesting that the cooling time from reconnection to plasma emission at temperature $\sim$1~MK ranges between 10 to 30 minutes. Employing observations of PRFLs in multiple EUV wavelengths, it is possible to derive time delays of plasma emissions characterizing temperatures from 10~MK to 1~MK with respect to the UV brightening (and other dynamic signatures from spectroscopic observations) at their foot-points. This helps establish the time sequence of heating and cooling for individual PRFLs, which will be pursued in our future work.


\begin{acknowledgments}

We thank the referee for insightful comments and suggestions that helped improve the paper.
This work was supported by NASA grant No. 80NSSC23K0414 and NSF grant No. AST-2407849. {\em SDO} is a mission of NASA's Living With a Star Program. 

\end{acknowledgments}



\bibliography{shear}{}

\begin{thebibliography}{}
\expandafter\ifx\csname natexlab\endcsname\relax\def\natexlab#1{#1}\fi
\providecommand{\url}[1]{\href{#1}{#1}}
\providecommand{\dodoi}[1]{doi:~\href{http://doi.org/#1}{\nolinkurl{#1}}}
\providecommand{\doeprint}[1]{\href{http://ascl.net/#1}{\nolinkurl{http://ascl.net/#1}}}
\providecommand{\doarXiv}[1]{\href{https://arxiv.org/abs/#1}{\nolinkurl{https://arxiv.org/abs/#1}}}

\bibitem[{J.~C. {Allred} {et~al.}(2015){Allred}, {Kowalski}, \&
  {Carlsson}}]{Allred2015}
{Allred}, J.~C., {Kowalski}, A.~F., \& {Carlsson}, M. 2015, \bibinfo{title}{{A
  Unified Computational Model for Solar and Stellar Flares},} \apj, 809, 104,
  \dodoi{10.1088/0004-637X/809/1/104}

\bibitem[{S.~K. {Antiochos} {et~al.}(2002){Antiochos}, {Karpen}, \&
  {DeVore}}]{Antiochos2002}
{Antiochos}, S.~K., {Karpen}, J.~T., \& {DeVore}, C.~R. 2002,
  \bibinfo{title}{{Coronal Magnetic Field Relaxation by Null-Point
  Reconnection},} \apj, 575, 578, \dodoi{10.1086/341193}

\bibitem[{E. {Antonucci} {et~al.}(1982){Antonucci}, {Gabriel}, {Acton},
  {Culhane}, {Doyle}, {Leibacher}, {Machado}, {Orwig}, \&
  {Rapley}}]{Antonucci1982}
{Antonucci}, E., {Gabriel}, A.~H., {Acton}, L.~W., {et~al.} 1982,
  \bibinfo{title}{{Impulsive Phase of Flares in Soft X-Ray Emission},}
  \solphys, 78, 107, \dodoi{10.1007/BF00151147}

\bibitem[{A. {Asai} {et~al.}(2004){Asai}, {Yokoyama}, {Shimojo}, {Masuda},
  {Kurokawa}, \& {Shibata}}]{Asai2004}
{Asai}, A., {Yokoyama}, T., {Shimojo}, M., {et~al.} 2004,
  \bibinfo{title}{{Flare Ribbon Expansion and Energy Release Rate},} \apj, 611,
  557, \dodoi{10.1086/422159}

\bibitem[{M.~J. {Aschwanden}(2010){Aschwanden}}]{Aschwanden2010}
{Aschwanden}, M.~J. 2010, \bibinfo{title}{{A Code for Automated Tracing of
  Coronal Loops Approaching Visual Perception},} \solphys, 262, 399,
  \dodoi{10.1007/s11207-010-9531-6}

\bibitem[{M.~J. {Aschwanden} \& D. {Alexander}(2001){Aschwanden} \&
  {Alexander}}]{Aschwanden2001}
{Aschwanden}, M.~J., \& {Alexander}, D. 2001, \bibinfo{title}{{Flare Plasma
  Cooling from 30 MK down to 1 MK modeled from Yohkoh, GOES, and TRACE
  observations during the Bastille Day Event (14 July 2000)},} \solphys, 204,
  91, \dodoi{10.1023/A:1014257826116}

\bibitem[{M.~J. {Aschwanden} \& H. {Peter}(2017){Aschwanden} \&
  {Peter}}]{Aschwanden2017}
{Aschwanden}, M.~J., \& {Peter}, H. 2017, \bibinfo{title}{{The Width
  Distribution of Loops and Strands in the Solar Corona{\textemdash}Are We
  Hitting Rock Bottom?},} \apj, 840, 4, \dodoi{10.3847/1538-4357/aa6b01}

\bibitem[{G. {Aulanier} {et~al.}(2012){Aulanier}, {Janvier}, \&
  {Schmieder}}]{Aulanier2012}
{Aulanier}, G., {Janvier}, M., \& {Schmieder}, B. 2012, \bibinfo{title}{{The
  standard flare model in three dimensions. I. Strong-to-weak shear transition
  in post-flare loops},} \aap, 543, A110, \dodoi{10.1051/0004-6361/201219311}

\bibitem[{M.~G. {Bobra} \& S. {Couvidat}(2015){Bobra} \&
  {Couvidat}}]{Bobra2015}
{Bobra}, M.~G., \& {Couvidat}, S. 2015, \bibinfo{title}{{Solar Flare Prediction
  Using SDO/HMI Vector Magnetic Field Data with a Machine-learning Algorithm},}
  \apj, 798, 135, \dodoi{10.1088/0004-637X/798/2/135}

\bibitem[{M.~G. {Bobra} {et~al.}(2014){Bobra}, {Sun}, {Hoeksema}, {Turmon},
  {Liu}, {Hayashi}, {Barnes}, \& {Leka}}]{Bobra2014}
{Bobra}, M.~G., {Sun}, X., {Hoeksema}, J.~T., {et~al.} 2014,
  \bibinfo{title}{{The Helioseismic and Magnetic Imager (HMI) Vector Magnetic
  Field Pipeline: SHARPs - Space-Weather HMI Active Region Patches},} \solphys,
  289, 3549, \dodoi{10.1007/s11207-014-0529-3}

\bibitem[{S.~A. {Bogachev} {et~al.}(2005){Bogachev}, {Somov}, {Kosugi}, \&
  {Sakao}}]{Bogachev2005}
{Bogachev}, S.~A., {Somov}, B.~V., {Kosugi}, T., \& {Sakao}, T. 2005,
  \bibinfo{title}{{The Motions of the Hard X-Ray Sources in Solar Flares:
  Images and Statistics},} \apj, 630, 561, \dodoi{10.1086/431918}

\bibitem[{H. {Carmichael}(1964){Carmichael}}]{Carmichael1964}
{Carmichael}, H. 1964, \bibinfo{title}{{A Process for Flares},} NASSP, 50, 451

\bibitem[{J.~T. {Dahlin} {et~al.}(2022){Dahlin}, {Antiochos}, {Qiu}, \&
  {DeVore}}]{Dahlin2022}
{Dahlin}, J.~T., {Antiochos}, S.~K., {Qiu}, J., \& {DeVore}, C.~R. 2022,
  \bibinfo{title}{{Variability of the Reconnection Guide Field in Solar
  Flares},} \apj, 932, 94, \dodoi{10.3847/1538-4357/ac6e3d}

\bibitem[{G.~H. {Fisher}(1987){Fisher}}]{Fisher1987}
{Fisher}, G.~H. 1987, \bibinfo{title}{{Explosive Evaporation in Solar Flares},}
  \apj, 317, 502, \dodoi{10.1086/165294}

\bibitem[{L. {Fletcher} \& H. {Hudson}(2001){Fletcher} \&
  {Hudson}}]{Fletcher2001}
{Fletcher}, L., \& {Hudson}, H. 2001, \bibinfo{title}{{The Magnetic Structure
  and Generation of EUV Flare Ribbons},} \solphys, 204, 69,
  \dodoi{10.1023/A:1014275821318}

\bibitem[{T.~G. {Forbes} \& L.~W. {Acton}(1996){Forbes} \&
  {Acton}}]{Forbes1996}
{Forbes}, T.~G., \& {Acton}, L.~W. 1996, \bibinfo{title}{{Reconnection and
  Field Line Shrinkage in Solar Flares},} \apj, 459, 330,
  \dodoi{10.1086/176896}

\bibitem[{P.~C. {Grigis} \& A.~O. {Benz}(2005){Grigis} \& {Benz}}]{Grigis2005}
{Grigis}, P.~C., \& {Benz}, A.~O. 2005, \bibinfo{title}{{The Evolution of
  Reconnection along an Arcade of Magnetic Loops},} \apjl, 625, L143,
  \dodoi{10.1086/431147}

\bibitem[{J. {Hinterreiter} {et~al.}(2018){Hinterreiter}, {Veronig},
  {Thalmann}, {Tschernitz}, \& {P{\"o}tzi}}]{Hinterreiter2018}
{Hinterreiter}, J., {Veronig}, A.~M., {Thalmann}, J.~K., {Tschernitz}, J., \&
  {P{\"o}tzi}, W. 2018, \bibinfo{title}{{Statistical Properties of Ribbon
  Evolution and Reconnection Electric Fields in Eruptive and Confined Flares},}
  \solphys, 293, 38, \dodoi{10.1007/s11207-018-1253-1}

\bibitem[{T. {Hirayama}(1974){Hirayama}}]{Hirayama1974}
{Hirayama}, T. 1974, \bibinfo{title}{{Theoretical Model of Flares and
  Prominences. I: Evaporating Flare Model},} \solphys, 34, 323,
  \dodoi{10.1007/BF00153671}

\bibitem[{A.~R. {Inglis} \& H.~R. {Gilbert}(2013){Inglis} \&
  {Gilbert}}]{Inglis2013}
{Inglis}, A.~R., \& {Gilbert}, H.~R. 2013, \bibinfo{title}{{Hard X-Ray and
  Ultraviolet Emission during the 2011 June 7 Solar Flare},} \apj, 777, 30,
  \dodoi{10.1088/0004-637X/777/1/30}

\bibitem[{H. {Isobe} {et~al.}(2002){Isobe}, {Shibata}, \&
  {Machida}}]{Isobe2002}
{Isobe}, H., {Shibata}, K., \& {Machida}, S. 2002, \bibinfo{title}{{``Dawn-dusk
  asymmetry'' in solar coronal arcade formations},} GeoRL, 29, 2014,
  \dodoi{10.1029/2001GL013816}

\bibitem[{M. {Janvier} {et~al.}(2013){Janvier}, {Aulanier}, {Pariat}, \&
  {D{\'e}moulin}}]{Janvier2013}
{Janvier}, M., {Aulanier}, G., {Pariat}, E., \& {D{\'e}moulin}, P. 2013,
  \bibinfo{title}{{The standard flare model in three dimensions. III.
  Slip-running reconnection properties},} \aap, 555, A77,
  \dodoi{10.1051/0004-6361/201321164}

\bibitem[{H. {Ji} {et~al.}(2006){Ji}, {Huang}, {Wang}, {Zhou}, {Li}, {Zhang},
  \& {Song}}]{Ji2006}
{Ji}, H., {Huang}, G., {Wang}, H., {et~al.} 2006, \bibinfo{title}{{Converging
  Motion of H{\ensuremath{\alpha}} Conjugate Kernels: The Signature of Fast
  Relaxation of a Sheared Magnetic Field},} \apjl, 636, L173,
  \dodoi{10.1086/500203}

\bibitem[{N.~C. {Joshi} {et~al.}(2017){Joshi}, {Sterling}, {Moore}, {Magara},
  \& {Moon}}]{Joshi2017}
{Joshi}, N.~C., {Sterling}, A.~C., {Moore}, R.~L., {Magara}, T., \& {Moon},
  Y.-J. 2017, \bibinfo{title}{{Onset of a Large Ejective Solar Eruption from a
  Typical Coronal-jet-base Field Configuration},} \apj, 845, 26,
  \dodoi{10.3847/1538-4357/aa7c1b}

\bibitem[{M.~D. {Kazachenko} {et~al.}(2022){Kazachenko}, {Lynch}, {Savcheva},
  {Sun}, \& {Welsch}}]{Kazachenko2022}
{Kazachenko}, M.~D., {Lynch}, B.~J., {Savcheva}, A., {Sun}, X., \& {Welsch},
  B.~T. 2022, \bibinfo{title}{{Toward Improved Understanding of Magnetic Fields
  Participating in Solar Flares: Statistical Analysis of Magnetic Fields within
  Flare Ribbons},} \apj, 926, 56, \dodoi{10.3847/1538-4357/ac3af3}

\bibitem[{M.~D. {Kazachenko} {et~al.}(2017){Kazachenko}, {Lynch}, {Welsch}, \&
  {Sun}}]{Kazachenko2017}
{Kazachenko}, M.~D., {Lynch}, B.~J., {Welsch}, B.~T., \& {Sun}, X. 2017,
  \bibinfo{title}{{A Database of Flare Ribbon Properties from the Solar
  Dynamics Observatory. I. Reconnection Flux},} \apj, 845, 49,
  \dodoi{10.3847/1538-4357/aa7ed6}

\bibitem[{G.~S. {Kerr} {et~al.}(2020){Kerr}, {Allred}, \& {Polito}}]{Kerr2020}
{Kerr}, G.~S., {Allred}, J.~C., \& {Polito}, V. 2020, \bibinfo{title}{{Solar
  Flare Arcade Modelling: Bridging the gap from 1D to 3D Simulations of
  Optically Thin Radiation},} arXiv e-prints, arXiv:2007.13856.
\newblock \doarXiv{2007.13856}

\bibitem[{T. {Kitahara} \& H. {Kurokawa}(1990){Kitahara} \&
  {Kurokawa}}]{Kitahara1990}
{Kitahara}, T., \& {Kurokawa}, H. 1990, \bibinfo{title}{{High-resolution
  observation and detailed photometry of a great H-alpha two-ribbon flare},}
  \solphys, 125, 321, \dodoi{10.1007/BF00158409}

\bibitem[{R.~A. {Kopp} \& G.~W. {Pneuman}(1976){Kopp} \& {Pneuman}}]{Kopp1976}
{Kopp}, R.~A., \& {Pneuman}, G.~W. 1976, \bibinfo{title}{{Magnetic reconnection
  in the corona and the loop prominence phenomenon.},} \solphys, 50, 85,
  \dodoi{10.1007/BF00206193}

\bibitem[{A.~F. {Kowalski} {et~al.}(2017){Kowalski}, {Allred}, {Daw}, {Cauzzi},
  \& {Carlsson}}]{Kowalski2017}
{Kowalski}, A.~F., {Allred}, J.~C., {Daw}, A., {Cauzzi}, G., \& {Carlsson}, M.
  2017, \bibinfo{title}{{The Atmospheric Response to High Nonthermal Electron
  Beam Fluxes in Solar Flares. I. Modeling the Brightest NUV Footpoints in the
  X1 Solar Flare of 2014 March 29},} \apj, 836, 12,
  \dodoi{10.3847/1538-4357/836/1/12}

\bibitem[{J.~R. {Lemen} {et~al.}(2012){Lemen}, {Title}, {Akin}, {Boerner},
  {Chou}, {Drake}, {Duncan}, {Edwards}, {Friedlaender}, {Heyman}, {Hurlburt},
  {Katz}, {Kushner}, {Levay}, {Lindgren}, {Mathur}, {McFeaters}, {Mitchell},
  {Rehse}, {Schrijver}, {Springer}, {Stern}, {Tarbell}, {Wuelser}, {Wolfson},
  {Yanari}, {Bookbinder}, {Cheimets}, {Caldwell}, {Deluca}, {Gates}, {Golub},
  {Park}, {Podgorski}, {Bush}, {Scherrer}, {Gummin}, {Smith}, {Auker},
  {Jerram}, {Pool}, {Soufli}, {Windt}, {Beardsley}, {Clapp}, {Lang}, \&
  {Waltham}}]{Lemen2012}
{Lemen}, J.~R., {Title}, A.~M., {Akin}, D.~J., {et~al.} 2012,
  \bibinfo{title}{{The Atmospheric Imaging Assembly (AIA) on the Solar Dynamics
  Observatory (SDO)},} \solphys, 275, 17, \dodoi{10.1007/s11207-011-9776-8}

\bibitem[{T. {Li} {et~al.}(2019){Li}, {Liu}, {Hou}, \& {Zhang}}]{Li2019}
{Li}, T., {Liu}, L., {Hou}, Y., \& {Zhang}, J. 2019, \bibinfo{title}{{Two Types
  of Confined Solar Flares},} \apj, 881, 151, \dodoi{10.3847/1538-4357/ab3121}

\bibitem[{C. {Liu} \& H. {Wang}(2009){Liu} \& {Wang}}]{Liu2009a}
{Liu}, C., \& {Wang}, H. 2009, \bibinfo{title}{{Reconnection Electric Field and
  Hardness of X-Ray Emission of Solar Flares},} \apjl, 696, L27,
  \dodoi{10.1088/0004-637X/696/1/L27}

\bibitem[{W. {Liu} {et~al.}(2009){Liu}, {Petrosian}, {Dennis}, \&
  {Holman}}]{Liu2009b}
{Liu}, W., {Petrosian}, V., {Dennis}, B.~R., \& {Holman}, G.~D. 2009,
  \bibinfo{title}{{Conjugate Hard X-Ray Footpoints in the 2003 October 29 X10
  Flare: Unshearing Motions, Correlations, and Asymmetries},} \apj, 693, 847,
  \dodoi{10.1088/0004-637X/693/1/847}

\bibitem[{Y. {Liu} {et~al.}(2023){Liu}, {Welsch}, {Valori}, {Georgoulis},
  {Guo}, {Pariat}, {Park}, \& {Thalmann}}]{Liu2023}
{Liu}, Y., {Welsch}, B.~T., {Valori}, G., {et~al.} 2023,
  \bibinfo{title}{{Changes of Magnetic Energy and Helicity in Solar Active
  Regions from Major Flares},} \apj, 942, 27, \dodoi{10.3847/1538-4357/aca3a6}

\bibitem[{D. {Longcope} {et~al.}(2018){Longcope}, {Unverferth}, {Klein},
  {McCarthy}, \& {Priest}}]{Longcope2018}
{Longcope}, D., {Unverferth}, J., {Klein}, C., {McCarthy}, M., \& {Priest}, E.
  2018, \bibinfo{title}{{Evidence for Downflows in the Narrow Plasma Sheet of
  2017 September 10 and Their Significance for Flare Reconnection},} \apj, 868,
  148, \dodoi{10.3847/1538-4357/aaeac4}

\bibitem[{D.~W. {Longcope}(2014){Longcope}}]{Longcope2014}
{Longcope}, D.~W. 2014, \bibinfo{title}{{A Simple Model of Chromospheric
  Evaporation and Condensation Driven Conductively in a Solar Flare},} \apj,
  795, 10, \dodoi{10.1088/0004-637X/795/1/10}

\bibitem[{D.~W. {Longcope} \& A. {Malanushenko}(2008){Longcope} \&
  {Malanushenko}}]{Longcope2008}
{Longcope}, D.~W., \& {Malanushenko}, A. 2008, \bibinfo{title}{{Defining and
  Calculating Self-Helicity in Coronal Magnetic Fields},} \apj, 674, 1130,
  \dodoi{10.1086/524011}

\bibitem[{A. {Malanushenko} {et~al.}(2009){Malanushenko}, {Longcope}, \&
  {McKenzie}}]{Malanushenko2009}
{Malanushenko}, A., {Longcope}, D.~W., \& {McKenzie}, D.~E. 2009,
  \bibinfo{title}{{Reconstructing the Local Twist of Coronal Magnetic Fields
  and the Three-Dimensional Shape of the Field Lines from Coronal Loops in
  Extreme-Ultraviolet and X-Ray Images},} \apj, 707, 1044,
  \dodoi{10.1088/0004-637X/707/2/1044}

\bibitem[{S. {Masuda} {et~al.}(1994){Masuda}, {Kosugi}, {Hara}, {Tsuneta}, \&
  {Ogawara}}]{Masuda1994}
{Masuda}, S., {Kosugi}, T., {Hara}, H., {Tsuneta}, S., \& {Ogawara}, Y. 1994,
  \bibinfo{title}{{A loop-top hard X-ray source in a compact solar flare as
  evidence for magnetic reconnection},} \nat, 371, 495,
  \dodoi{10.1038/371495a0}

\bibitem[{D. {Nandy} {et~al.}(2003){Nandy}, {Hahn}, {Canfield}, \&
  {Longcope}}]{Nandy2003}
{Nandy}, D., {Hahn}, M., {Canfield}, R.~C., \& {Longcope}, D.~W. 2003,
  \bibinfo{title}{{Detection of a Taylor-like Plasma Relaxation Process in the
  Sun},} \apjl, 597, L73, \dodoi{10.1086/379815}

\bibitem[{S. {Patsourakos} {et~al.}(2020){Patsourakos}, {Vourlidas},
  {T{\"o}r{\"o}k}, {Kliem}, {Antiochos}, {Archontis}, {Aulanier}, {Cheng},
  {Chintzoglou}, {Georgoulis}, {Green}, {Leake}, {Moore}, {Nindos}, {Syntelis},
  {Yardley}, {Yurchyshyn}, \& {Zhang}}]{Patsourakos20}
{Patsourakos}, S., {Vourlidas}, A., {T{\"o}r{\"o}k}, T., {et~al.} 2020,
  \bibinfo{title}{{Decoding the Pre-Eruptive Magnetic Field Configurations of
  Coronal Mass Ejections},} \ssr, 216, 131, \dodoi{10.1007/s11214-020-00757-9}

\bibitem[{W.~D. {Pesnell} {et~al.}(2012){Pesnell}, {Thompson}, \&
  {Chamberlin}}]{Pesnell2012}
{Pesnell}, W.~D., {Thompson}, B.~J., \& {Chamberlin}, P.~C. 2012,
  \bibinfo{title}{{The Solar Dynamics Observatory (SDO)},} \solphys, 275, 3,
  \dodoi{10.1007/s11207-011-9841-3}

\bibitem[{A. {Prasad} {et~al.}(2023){Prasad}, {Kumar}, {Sterling}, {Moore},
  {Aulanier}, {Bhattacharyya}, \& {Hu}}]{Prasad2023}
{Prasad}, A., {Kumar}, S., {Sterling}, A.~C., {et~al.} 2023,
  \bibinfo{title}{{Formation of an observed eruptive flux rope above the torus
  instability threshold through tether-cutting magnetic reconnection},} \aap,
  677, A43, \dodoi{10.1051/0004-6361/202346267}

\bibitem[{J. {Qiu}(2009){Qiu}}]{Qiu2009}
{Qiu}, J. 2009, \bibinfo{title}{{Observational Analysis of Magnetic
  Reconnection Sequence},} \apj, 692, 1110,
  \dodoi{10.1088/0004-637X/692/2/1110}

\bibitem[{J. {Qiu} \& J. {Cheng}(2022){Qiu} \& {Cheng}}]{Qiu2022}
{Qiu}, J., \& {Cheng}, J. 2022, \bibinfo{title}{{Properties and Energetics of
  Magnetic Reconnection: I. Evolution of Flare Ribbons},} \solphys, 297, 80,
  \dodoi{10.1007/s11207-022-02003-7}

\bibitem[{J. {Qiu} \& R. {Fleming}(2025){Qiu} \& {Fleming}}]{Qiu2025}
{Qiu}, J., \& {Fleming}, R. 2025, \bibinfo{title}{{Quantifying Chromosphere
  Response to Flare Energy Release Using AIA Observations in
  1600\raisebox{-0.5ex}\textasciitilde{\r{A}} and
  304\raisebox{-0.5ex}\textasciitilde{\r{A}} Passbands},} arXiv e-prints,
  arXiv:2505.13728, \dodoi{10.48550/arXiv.2505.13728}

\bibitem[{J. {Qiu} {et~al.}(2002){Qiu}, {Lee}, {Gary}, \& {Wang}}]{Qiu2002}
{Qiu}, J., {Lee}, J., {Gary}, D.~E., \& {Wang}, H. 2002,
  \bibinfo{title}{{Motion of Flare Footpoint Emission and Inferred Electric
  Field in Reconnecting Current Sheets},} \apj, 565, 1335,
  \dodoi{10.1086/324706}

\bibitem[{J. {Qiu} {et~al.}(2010){Qiu}, {Liu}, {Hill}, \&
  {Kazachenko}}]{Qiu2010}
{Qiu}, J., {Liu}, W., {Hill}, N., \& {Kazachenko}, M. 2010,
  \bibinfo{title}{{Reconnection and Energetics in Two-ribbon Flares: A Revisit
  of the Bastille-day Flare},} \apj, 725, 319,
  \dodoi{10.1088/0004-637X/725/1/319}

\bibitem[{J. {Qiu} \& D.~W. {Longcope}(2016){Qiu} \& {Longcope}}]{Qiu2016}
{Qiu}, J., \& {Longcope}, D.~W. 2016, \bibinfo{title}{{Long Duration Flare
  Emission: Impulsive Heating or Gradual Heating?},} \apj, 820, 14,
  \dodoi{10.3847/0004-637X/820/1/14}

\bibitem[{J. {Qiu} {et~al.}(2017){Qiu}, {Longcope}, {Cassak}, \&
  {Priest}}]{Qiu2017}
{Qiu}, J., {Longcope}, D.~W., {Cassak}, P.~A., \& {Priest}, E.~R. 2017,
  \bibinfo{title}{{Elongation of Flare Ribbons},} \apj, 838, 17,
  \dodoi{10.3847/1538-4357/aa6341}

\bibitem[{J. {Qiu} {et~al.}(2023){Qiu}, {Alaoui}, {Antiochos}, {Dahlin},
  {Swisdak}, {Drake}, {Robison}, {DeVore}, \& {Uritsky}}]{Qiu2023}
{Qiu}, J., {Alaoui}, M., {Antiochos}, S.~K., {et~al.} 2023,
  \bibinfo{title}{{The Role of Magnetic Shear in Reconnection-driven Flare
  Energy Release},} \apj, 955, 34, \dodoi{10.3847/1538-4357/acebeb}

\bibitem[{J.~W. {Reep} {et~al.}(2020){Reep}, {Warren}, {Moore}, {Suarez}, \&
  {Hayes}}]{Reep2020}
{Reep}, J.~W., {Warren}, H.~P., {Moore}, C.~S., {Suarez}, C., \& {Hayes}, L.~A.
  2020, \bibinfo{title}{{Simulating Solar Flare Irradiance with Multithreaded
  Models of Flare Arcades},} \apj, 895, 30, \dodoi{10.3847/1538-4357/ab89a0}

\bibitem[{T. {Sakao}(1994){Sakao}}]{Sakao1994}
{Sakao}, T. 1994, PhD thesis, Univ.\ Tokyo

\bibitem[{J. {Schou} {et~al.}(2012){Schou}, {Scherrer}, {Bush}, {Wachter},
  {Couvidat}, {Rabello-Soares}, {Bogart}, {Hoeksema}, {Liu}, {Duvall}, {Akin},
  {Allard}, {Miles}, {Rairden}, {Shine}, {Tarbell}, {Title}, {Wolfson},
  {Elmore}, {Norton}, \& {Tomczyk}}]{Schou2012}
{Schou}, J., {Scherrer}, P.~H., {Bush}, R.~I., {et~al.} 2012,
  \bibinfo{title}{{Design and Ground Calibration of the Helioseismic and
  Magnetic Imager (HMI) Instrument on the Solar Dynamics Observatory (SDO)},}
  \solphys, 275, 229, \dodoi{10.1007/s11207-011-9842-2}

\bibitem[{C.~J. {Schrijver} {et~al.}(2008){Schrijver}, {DeRosa}, {Metcalf},
  {Barnes}, {Lites}, {Tarbell}, {McTiernan}, {Valori}, {Wiegelmann},
  {Wheatland}, {Amari}, {Aulanier}, {D{\'e}moulin}, {Fuhrmann}, {Kusano},
  {R{\'e}gnier}, \& {Thalmann}}]{Schrijver2008}
{Schrijver}, C.~J., {DeRosa}, M.~L., {Metcalf}, T., {et~al.} 2008,
  \bibinfo{title}{{Nonlinear Force-free Field Modeling of a Solar Active Region
  around the Time of a Major Flare and Coronal Mass Ejection},} \apj, 675,
  1637, \dodoi{10.1086/527413}

\bibitem[{P.~A. {Sturrock}(1966){Sturrock}}]{Sturrock1966}
{Sturrock}, P.~A. 1966, \bibinfo{title}{{Model of the High-Energy Phase of
  Solar Flares},} \nat, 211, 695, \dodoi{10.1038/211695a0}

\bibitem[{Y. {Su} {et~al.}(2007){Su}, {Golub}, \& {Van Ballegooijen}}]{Su2007}
{Su}, Y., {Golub}, L., \& {Van Ballegooijen}, A.~A. 2007, \bibinfo{title}{{A
  Statistical Study of Shear Motion of the Footpoints in Two-Ribbon Flares},}
  \apj, 655, 606, \dodoi{10.1086/510065}

\bibitem[{Y.~N. {Su} {et~al.}(2006){Su}, {Golub}, {van Ballegooijen}, \&
  {Gros}}]{Su2006}
{Su}, Y.~N., {Golub}, L., {van Ballegooijen}, A.~A., \& {Gros}, M. 2006,
  \bibinfo{title}{{Analysis of Magnetic Shear in An X17 Solar Flare on October
  28, 2003},} \solphys, 236, 325, \dodoi{10.1007/s11207-006-0039-z}

\bibitem[{X. {Sun} {et~al.}(2012){Sun}, {Hoeksema}, {Liu}, {Wiegelmann},
  {Hayashi}, {Chen}, \& {Thalmann}}]{Sun2012}
{Sun}, X., {Hoeksema}, J.~T., {Liu}, Y., {et~al.} 2012,
  \bibinfo{title}{{Evolution of Magnetic Field and Energy in a Major Eruptive
  Active Region Based on SDO/HMI Observation},} \apj, 748, 77,
  \dodoi{10.1088/0004-637X/748/2/77}

\bibitem[{Z. {Svestka}(1980){Svestka}}]{Svestka1980}
{Svestka}, Z. 1980, \bibinfo{title}{{Activated Solar Filaments and Flares},}
  RSPTA, 297, 575, \dodoi{10.1098/rsta.1980.0233}

\bibitem[{M. {Temmer} {et~al.}(2007){Temmer}, {Veronig}, {Vr{\v s}nak}, \&
  {Miklenic}}]{Temmer2007}
{Temmer}, M., {Veronig}, A.~M., {Vr{\v s}nak}, B., \& {Miklenic}, C. 2007,
  \bibinfo{title}{{Energy Release Rates along H{$\alpha$} Flare Ribbons and the
  Location of Hard X-Ray Sources},} \apj, 654, 665, \dodoi{10.1086/509634}

\bibitem[{H.~P. {Warren}(2006){Warren}}]{Warren2006}
{Warren}, H.~P. 2006, \bibinfo{title}{{Multithread Hydrodynamic Modeling of a
  Solar Flare},} \apj, 637, 522, \dodoi{10.1086/497904}

\bibitem[{H.~P. {Warren} {et~al.}(2018){Warren}, {Crump}, {Ugarte-Urra}, {Sun},
  {Aschwanden}, \& {Wiegelmann}}]{Warren2018}
{Warren}, H.~P., {Crump}, N.~A., {Ugarte-Urra}, I., {et~al.} 2018,
  \bibinfo{title}{{Toward a Quantitative Comparison of Magnetic Field
  Extrapolations and Observed Coronal Loops},} \apj, 860, 46,
  \dodoi{10.3847/1538-4357/aac20b}

\bibitem[{T. {Wiegelmann} \& T. {Sakurai}(2021){Wiegelmann} \&
  {Sakurai}}]{Wiegelmann2021}
{Wiegelmann}, T., \& {Sakurai}, T. 2021, \bibinfo{title}{{Solar force-free
  magnetic fields},} Living Reviews in Solar Physics, 18, 1,
  \dodoi{10.1007/s41116-020-00027-4}

\bibitem[{Y.-H. {Yang} {et~al.}(2011){Yang}, {Cheng}, {Krucker}, \&
  {Hsieh}}]{Yang2011}
{Yang}, Y.-H., {Cheng}, C.~Z., {Krucker}, S., \& {Hsieh}, M.-S. 2011,
  \bibinfo{title}{{Estimation of the Reconnection Electric Field in the 2003
  October 29 X10 Flare},} \apj, 732, 15, \dodoi{10.1088/0004-637X/732/1/15}

\bibitem[{Y.-H. {Yang} {et~al.}(2009){Yang}, {Cheng}, {Krucker}, {Lin}, \&
  {Ip}}]{Yang2009}
{Yang}, Y.-H., {Cheng}, C.~Z., {Krucker}, S., {Lin}, R.~P., \& {Ip}, W.~H.
  2009, \bibinfo{title}{{A Statistical Study of Hard X-Ray Footpoint Motions in
  Large Solar Flares},} \apj, 693, 132, \dodoi{10.1088/0004-637X/693/1/132}

\bibitem[{I.~V. {Zimovets} {et~al.}(2020){Zimovets}, {Sharykin}, \&
  {Gan}}]{Zimovets2020}
{Zimovets}, I.~V., {Sharykin}, I.~N., \& {Gan}, W.~Q. 2020,
  \bibinfo{title}{{Relationships between Photospheric Vertical Electric
  Currents and Hard X-Ray Sources in Solar Flares: Statistical Study},} \apj,
  891, 138, \dodoi{10.3847/1538-4357/ab75be}

\end{thebibliography}
\bibliographystyle{aasjournal}

\allauthors


\end{document}